\documentclass[a4paper]{aa}
\usepackage{amssymb}
\usepackage{graphicx}
\usepackage{natbib}
\usepackage{hyperref}
\usepackage{soul}
\usepackage{color}
\usepackage{enumerate}

\begin{document}

\newcommand{\CaAX}{$A^2\Pi_r - X^2\Sigma^{+}$} 
\newcommand{\CaBX}{$B/B^\prime 2\Sigma^{+} - X^2\Sigma^{+}$} 
\def\AX{$A~^2\Pi~-~X~^2\Sigma^+$~}
\def\gb{G~224-58~B~}
\def\ga{G~224-58~A~}
\def\gab{G~224-58~AB}
\def\Tef{$T_{\rm eff}$~}
\def\AE{$F_{\rm a}$~}
\def\Vsini{$v$ sin $i$~}
\def\Vm{$V_{\rm m}$~}
\def\EE{$E^{''}$~}
\def\kmps{kms$^{\rm -1}$~}
\def\um{$\mu$m~}
\def\logg{$\log~g$~}

\title{Probing M subdwarf metallicity with an esdK5+esdM5.5 binary}


\author{Ya.~V. Pavlenko \inst{1,2}, Z.~H. Zhang\inst{3,4},
 M.~C. G\'alvez-Ortiz\inst{5,1}, I.~O. Kushniruk\inst{6},
H.~R.~A. Jones\inst{1}
}
\institute{Centre for Astrophysics Research, University of Hertfordshire,
College Lane, Hatfield, Hertfordshire AL10 9AB, UK
\and Main Astronomical Observatory, Academy of Sciences of the Ukraine, Golosiiv
     Woods, Kyiv-127, 03680 Ukraine
\and Instituto de Astrof\' isika  de Canarias, E-38205 La Laguna, Tenerife, Spain
\and Universidad de La Laguna, Dept. Astrof\'i sica, E-38206.
La Laguna, Tenerife, Spain
\and  Centro de Astrobiolog\' ia (CSIC-INTA), Ctra. Ajalvir km 4, E-28850 
Torre{\'j}on de Ardoz, Madrid, Spain
\and Taras Shevchenko National University of Kyiv, 
      60 Volodymyrska Str., Kyiv, 01033 Ukraine
}

\offprints{Ya. V. Pavlenko}
\mail{email2yp@gmail.com}

\date{}

\authorrunning{Pavlenko et al.}
\titlerunning{The study of the esdK5+esdM5.5 binary}

\abstract{
We present a spectral analysis of the binary G 224-58 AB 
 that consists
 of the coolest M extreme subdwarf (esdM5.5) and a brighter primary (esdK5). 
This binary may serve as a benchmark for 
 metallicity measurement calibrations and as a test-bed for atmospheric 
and evolutionary models for esdM objects.
}{
We perform the analysis of optical and infrared spectra of both components 
to determine their parameters.
}{
We determine abundances primarily using high resolution optical spectra of the primary.
Other parameters 
were determined from the fits of synthetic spectra computed 
with these abundances to the observed spectra
from 0.4 to 2.5 microns for both components.
}{
We determine \Tef =4625 $\pm$ 100 K, 
\logg = 4.5 $\pm$ 0.5 for the A component and \Tef = 3200 $\pm$ 100 K, 
\logg = 5.0 $\pm$ 0.5, for  
the B component. We obtained abundances of [Mg/H]=$-$1.51$\pm$0.08, 
[Ca/H]=$-$1.39$\pm$0.03, [Ti/H]=$-$1.37$\pm$0.03 for alpha group elements 
and [CrH]=$-$1.88$\pm$0.07, [Mn/H]=$-$1.96$\pm$0.06, [Fe/H]=$-$1.92$\pm$0.02, 
[Ni/H]=$-$1.81$\pm$0.05 and [Ba/H]=$-$1.87$\pm$0.11 for iron group elements 
from fits to the spectral lines observed in the optical and infrared 
spectral regions of the primary. We find consistent abundances with 
fits to the secondary albeit at lower signal-to-noise. 
}{
 Abundances of different elements  in \ga and \gb atmospheres cannot be 
described  by one metallicity parameter.
The offset of  $\sim$ 0.4 dex  between the 
abundances derived from alpha element and iron group elements 
corresponds with our expectation for 
metal-deficient stars.
We thus clarify that  some indices used to date 
to measure metallicities for  establishing  esdM stars based on CaH, MgH and 
TiO band system strength ratios in the optical  
and H$_2$O in the infrared relate to abundances of 
alpha-element group rather   
than to iron peak elements. For metal deficient M dwarfs with 
[Fe/H] < -1.0, this provides a ready explanation 
for apparently inconsistent "metallicities" derived using different methods.}


\maketitle 

\section{Introduction}

Very low-mass stars or M dwarfs are the most numerous and 
longest-lived stars in our Milky Way. They are targeted for the study 
of the Galactic structure and searches of habitable exoplanets by a 
 large number of projects.  

The precise metallicity measurement of M dwarfs has become a popular 
field  \citep[e.g.,][]{roja10, terr12, oneh12, neve13, newt14}. However, the data quality of observations and models make it 
useful to analyse M dwarfs in binary systems with 
F-, G- or K-type dwarfs in order to ensure robust calibration of metallicity features in their 
spectra. These M dwarfs not only share the same age and distance with their 
FGK companions. They also share chemical abundance which have been  well calibrated 
for FGK dwarfs. Such studies are limited to the M dwarfs in the Galactic disk 
where there exist enough wide FGK+M dwarf binary systems for calibration.

Sub classifications for low metallicity M dwarfs or M subdwarfs  \citep{gizi97,lepi07} are widely in use. 
Precise metallicity measurements of M subdwarfs have been recently 
conducted by \cite{wool09} and \cite{rajp14}. 
  They both used high resolution spectra of subdwarfs to measure metallicities and calibrate
 a relation between [Fe/H] and  molecular band strength indices from low-resolution spectra.
  This calibration makes  possible to estimate the metallicity
 of a large sample of subdwarfs without the observational cost
 (some times impossible) of high resolution spectroscopy of these faint objects,
 to enable meaningful population analysis.
 \cite{wool09} measured [Fe/H] from Fe I lines using NextGen models
 to fit the spectra of 12 low-metallicity main sequence M stars. 
 Adding these objects to previous compilations \citep{wool06, wool09}
 they established the relationship between the $\zeta_{\rm TiO/CaH}$ parameter and metallicity
 for the objects with \Tef of 3500-4000 K, and  
metallicities over $\sim$-2. 
 \cite{rajp14} used BT-Settl models to fit the spectra of
 three late-K subdwarfs and 18 M subdwarfs (covering the M0-M9.5 range),
 reaching the lowest value of [Fe/H]=-1.7.
 These works  mainly use the spectra of single M subdwarfs,
 without the support of the measurements of any binary FGK companion.
 It is worth noting, that \cite{wool06} studied M dwarfs with a
G or warm K companion and M dwarfs
in the Hyades cluster as part of their sample.
Furthermore, \cite{bean06} reported tests
of metallicity measurements using M dwarfs in binaries
with warmer companions.

G 224-58 AB is a esdK5+esdM5.5 wide binary system with 93 arcsec separation 
identified by motion, colour and spectra in  \cite{zhan13}. \gb is the coolest M extreme subdwarf with a bright primary. 
In this paper, we study 
\ga through the analysis of high resolution observed and synthetic spectra. 
 This provides by association a more precise metallicity 
 of \gb\, a very cool extreme subdwarf that will
 serve as benchmark for calibration of metallicity 
 measurement and as a test-bed for atmospheric and 
 evolutionary models of such metal-poor low-mass objects.
 In Section \ref{_obs} we present 
 the observational data used in our analysis. 
Procedure and results of our analysis are given in Sections  \ref{_ga} and 
\ref{_gb}, respectively. Our main results are discussed in section \ref{_dc}.

\section{Observations \label{_obs}} 

\subsection{FIES optical spectrum of \ga}
High resolution spectrum of \ga were taken on 2013 June 19  
with the the high-resolution \citep[FIES;][]{telt14}) on the Nordic Optical Telescope.
The CCD13, EEV42-40 detector (2k$\times$2k) was used in medium resolution mode (R=46000) 
covering the 3630 - 7260 \AA \  interval in 80 orders.
Signal-to-noise is $\sim$35 at H$\alpha$. In general, this value is a bit low
for classical abundance analysis using the curve-of growth procedure. In our case only 
well defined and fitted parts of spectral lines are used. Another problem is the 
determination of the continuum level.

Data were reduced using the standard reduction procedures in 
  IRAF\footnote{IRAF is distributed by the National Optical Observatory,
 which is operated by the Association of Universities for Research in
 Astronomy, Inc., under contract with the National Science Foundation.}.
 We used the {\sc echelle} package for bias subtraction, 
 flat-field division, scatter correction, extraction of the spectra, 
 and wavelength calibration with Th-Ar lamps. 
 Since FIES have a dedicated data reduction pipeline, 
 FIESTool\footnote{http://www.not.iac.es/instruments/fies/fiestool/FIEStool.html},
  we compared our reductions with those of the pipeline 
 and found them to be equivalent.

\subsection{SDSS optical spectrum of \gb}
As in \cite{zhan13}, we used the
 SDSS J151650.33+605305.4 (G 224-58 B) spectrum from 
the Science Archive Server (SAS)\footnote{http://dr10.sdss3.org/}
that contains 
 the Tenth SDSS Data Release (DR10) of Sloan Digital Sky Survey \citep[SDSS;][]{york00}.
 The data is provided as final spectrum product from the pipeline.
 The SDSS spectroscopic pipelines, extract one dimensional spectra from 
 the raw exposures produced by the spectrographs, calibrate them in 
 wavelength and flux, combine the red and blue halves of the spectra, measure 
 features in these spectra,
 measure redshifts from these features, and classify the objects
 as galaxies, stars, or quasars
 (see http://www.sdss3.org/dr10/spectro/pipeline.php for more details).

\subsection{LIRIS near infrared spectra of \gab}
Medium resolution near infrared (NIR) spectra of G 224-58 AB were obtained 
 with the Long-slit Intermediate Resolution Infrared Spectrograph 
 \citep[LIRIS][]{manc98} at the William Herschel Telescope on 2015 May 3 and 4. 
 We used hr\_j, hr\_h and hr\_k grisms for 
 the observation, which provide a resolving power of 2500.  
 The data were reduced with the IRAF LIRISDR package which is supported by 
 Jose Antonio Acosta Pulido at the 
IAC\footnote{http://www.iac.es/galeria\-/jap/lirisdr\-/LIRIS\_\-DATA\_RE\-DUCTION.html}.
The total integration times used on \ga were 600s divided into ten 
60s exposures using the technique of ABBA exposures to remove sky background in  $J, H$ and $K$ bands. Three 
of the $K$ band spectra of \ga were taken on the gap between detectors 
and rejected for final combination. Thus the total integration time for 
final $K$ band spectrum of \ga is 420s. The total integration times used 
on \gb were 18 $\times$ 120s = 3600s in $J$ and $H$ bands and 
20 $\times$ 120s = 4000s in $K$ band.

\section{\ga spectrum analysis\label{_ga}}
\subsection{The procedure of the \ga\ abundance analysis. \label{_pa}} 

To determine the physical parameters of \ga we followed the procedure 
described in \cite{pavl12}. 
We determined abundances in the \ga atmosphere  iteratively. Namely, after the
 determination of abundances,  the model atmosphere was recomputed to 
 account for the changes to these abundances. The 1D model
atmospheres were computed by SAM12 program \citep{pavl03}. 
Absorption of atomic and molecular lines was accounted for by opacity sampling.
Spectroscopic data for absorption lines of atoms and molecules were taken
from the VALD2 \citep{kupk99} and Kurucz database \citep{kuru93}, 
respectively. 
The shape of each atomic line is determined
using a Voigt function and all damping constants are taken from line
databases, or computed using Unsold's approach (Unsold 1954).
For the complete description of routines  \citep[see][]{pavl12}.

Synthetic spectra are calculated with the WITA6 program \citep{pavl97},
using the same approximations and opacities as SAM12.
To compute the synthetic spectra we used the same line lists 
as for the model atmosphere computation. 
A Wavelength step of $\Delta\lambda$ = 0.025 \AA\ is employed in the
synthetic spectra computations. The synthetic
spectra that are computed across the selected spectral region are convolved
with profiles that match the instrumental broadening and that take into account
rotational broadening following the procedure described by \cite{gray76}.

\begin{figure}
\centering
\includegraphics[width=85mm, angle=0]{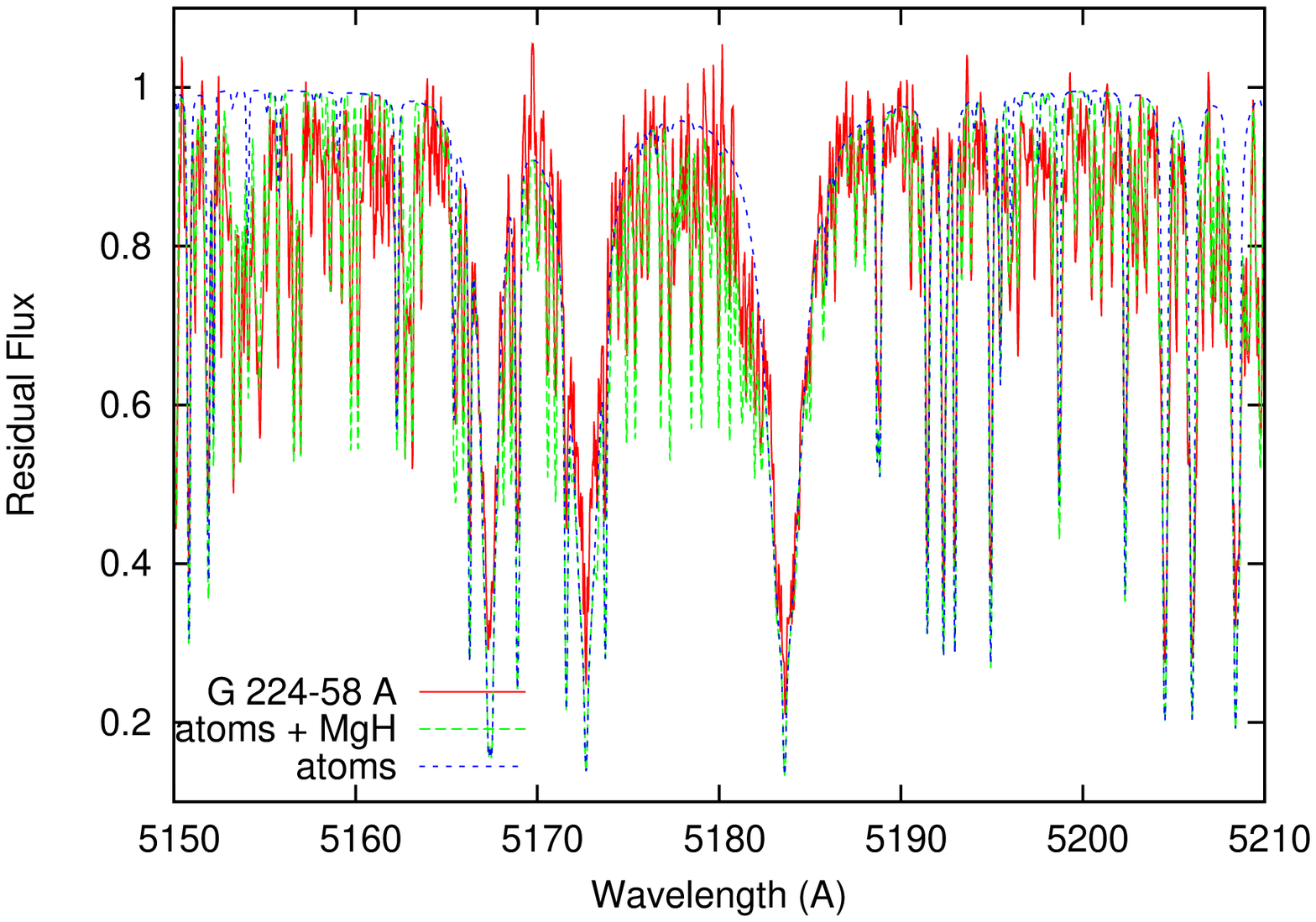}
\includegraphics[width=85mm, angle=0]{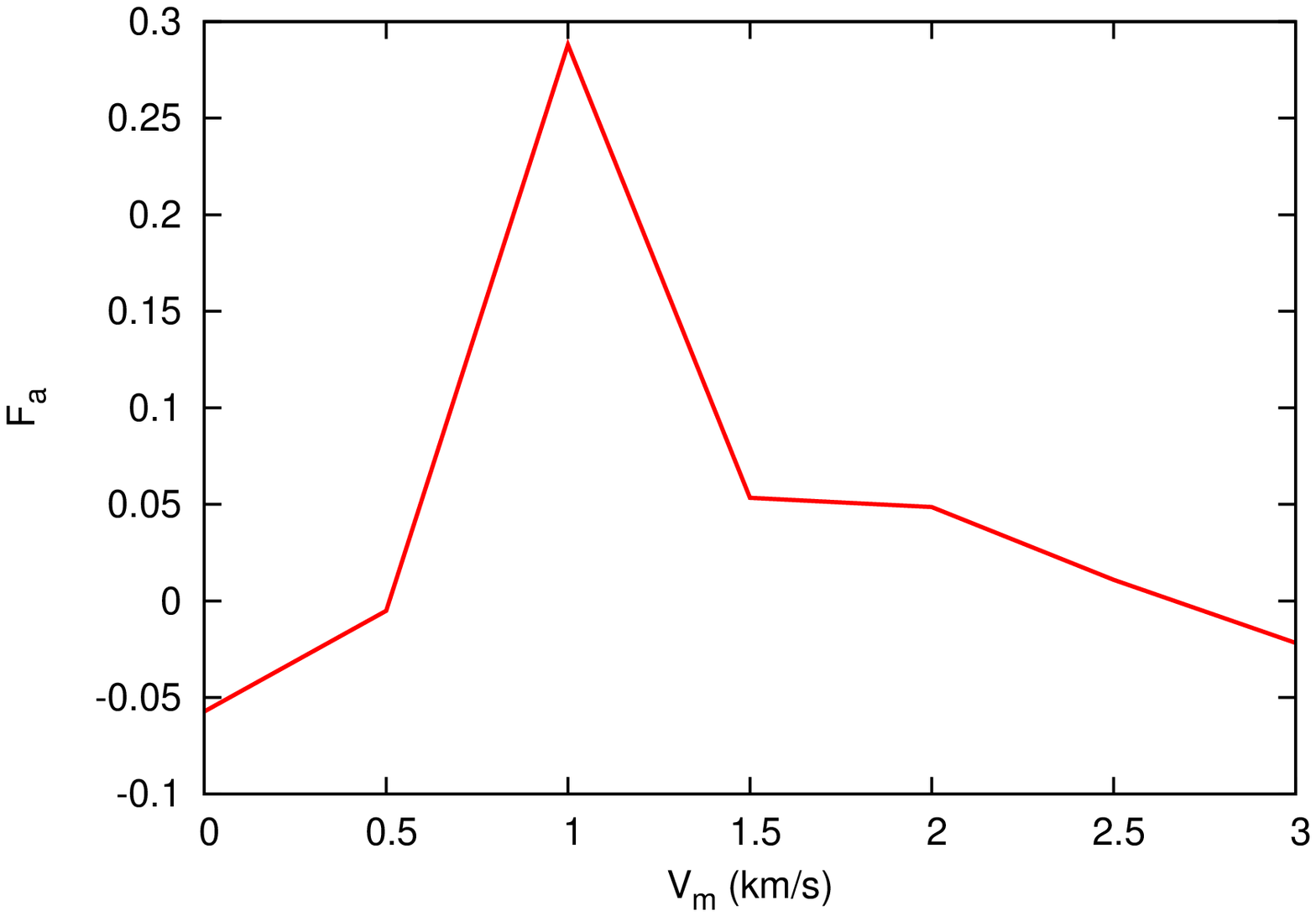}
\includegraphics[width=85mm, angle=0]{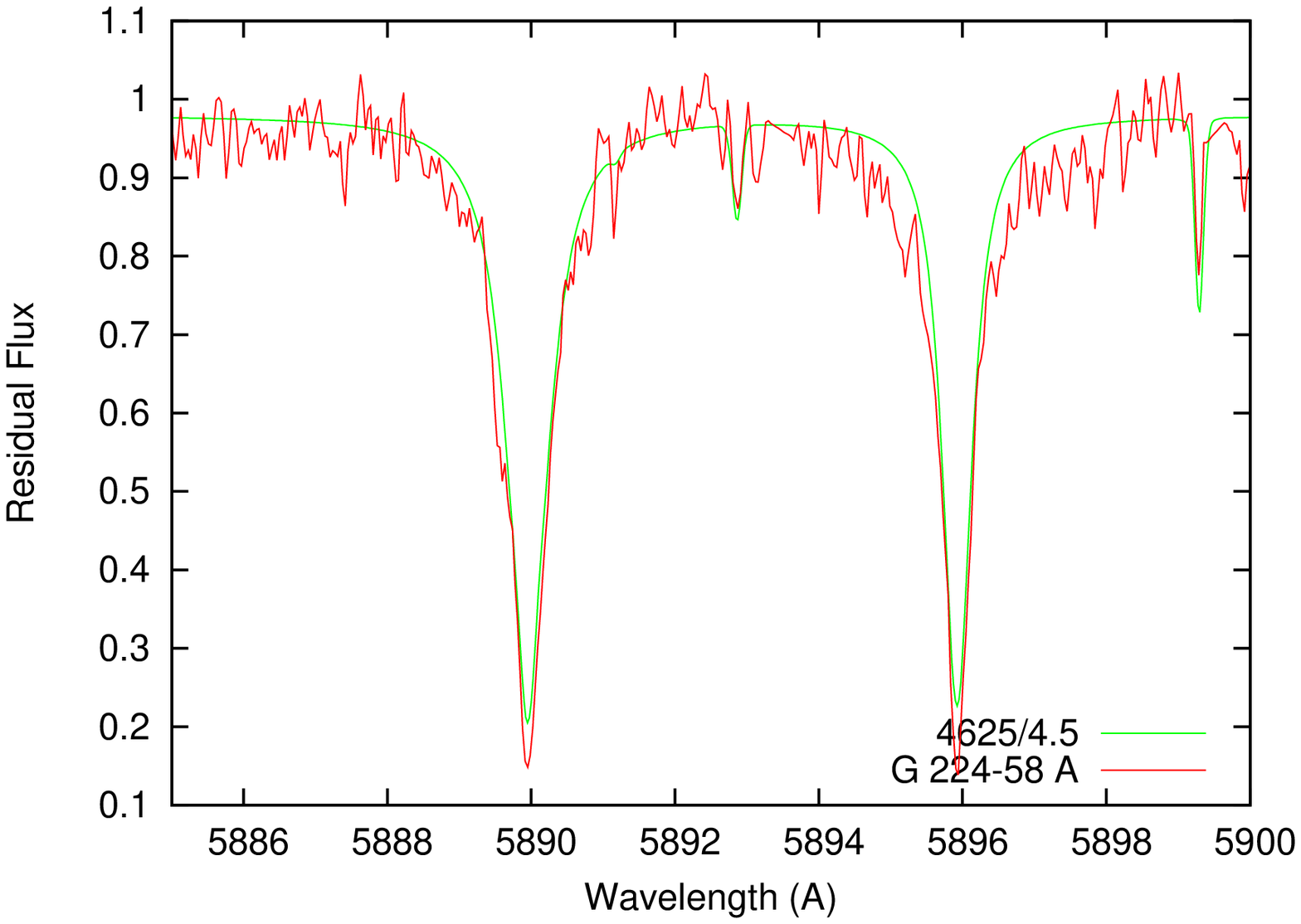}
\caption{Top: fits of the observed in spectrum \ga 
Mg I line triplet and MgH lines Labels 'atoms' and 'atoms+MgH' 
mark synthetic spectra computed 
with account of atomic lines and atomic lines + MgH, 
respectively. 
Middle: dependence of \AE on \Vm for Fe lines, 
model atmosphere 4625/4.5. Bottom: fits to Na I resonance doublet.
}
\label{_vt}
\end{figure}

We used the Sun as a template star in this study 
 applying our technique to determine solar abundances in order to give reliability to our abundances for \ga.
 We used an observed solar spectrum with resolution R $>$ 100,000 from \cite{kuru84}.
 The abundances are determined for a list of pre-selected lines which are fitted well in the solar spectrum and
 for the Sun we adopt
T$_{eff}$/$log$ g/[Fe/H]=5777/4.44/0.00 \citep[see][]{pavl12}). 
The abundances we obtained are in 
 good agreement with known classical results, e.g.
 \cite{_AG89, _GK89} within  $\pm$ 0.1 dex. 
 Further details of the procedure are given in  \cite{kush14}.

\subsection{Modelling \ga optical spectrum \label{_optA}}

From the literature, we expect  
that the effective temperature of \ga is in the 4500 - 4750 K range and 
that the atmosphere is metal deficient.
We carried out our abundance determination of 
the star applying the
procedure described in the section \ref{_pa} for model
atmospheres of \Tef = 4500 and 4750 K, log g = 4.5 and 5.0. These computations
provide rather similar results, i.e [Fe/H] $\sim$ -2.0. We probed these
solutions with analysis of the spectral region where strong lines of Mg I
and the \AX band system of MgH molecule
are located, see the upper panel of Fig. \ref{_vt}. 
These features in the synthetic spectra show a  
notable dependence on the gravity chosen \citep{kush13}.
Additionally we determined the titanium abundance log N(Ti) using the lines of
different titanium ions, i.e. Ti I and Ti  II. We found formally 
similar results of abundance determinations for the cases of 4500/4.5 and 4750/5.0.
However, MgH lines were too strong in the case of 4500/4.5 and too weak in the case of 
4750/5.0, and abundances obtained from the fits to Ti I and Ti II lines 
disagree by more than 0.1 dex with each other, see Section \ref{_tii}.



We checked  \Tef = 4625 K, and log g = 4.5--5.0 models.
Since 4625/4.5  model matches better both the MgH lines and the
minimal disagreement between Ti I and Ti II abundances (see section \ref{_tii}), 
 we choose it for our further analysis.

\subsubsection{Microturbulent velocity \Vm}

 The procedure of microturbulent velocity determination differs from that described in 
\cite{pavl12} as follows.
\begin{itemize} 

\item We determine a small grid of $N_v$  microturbulent velocities 
$V_m = V_0+i*0.5$, $i$ changes from 1 to 7. We adopt $V_0$ = 0 km/s and so cover a grid of 
plausible values for \Vm in the atmospheres.

\item For every microturbulent velocity from our grid we obtained the iron abundances 
$log N_k(Fe), k = 1, \dotsc  L,$ here $L$ is the number of lines in our pre-selected list, 
and determined the central residual fluxes of fitted lines $r_{\lambda_0^k}$. 

\item Using $\chi^2$ fitting, we approximate the dependence of log N(Fe) vs. 
$r_{\lambda_0}$ by the linear formula
$log N(Fe) = A$ + \AE * $(r_{\lambda_0})$, here $A$, \AE are constants which depend on \Vm.
We are interested in the case \AE = 0, when all lines provide the same abundances.

\end{itemize}

 A similar approach for \Vm determination from the 
equivalent width vs. abundance analysis was used by R. Kurucz in 
his WIDTH9 code. However, our approach is more flexible in the sense that we
can use  \Vm determination even for blended lines.

In the middle panel of  Fig. \ref{_vt} we see \AE = 0 at least at \Vm =0.5 and 2.5 km/s.
Corresponding abundances of iron obtained from the fits
 to profiles of 44 lines of Fe I in the spectrum of the 
primary are -6.290 $\pm$  0.021 and -6.478 $\pm$  
0.026, respectively. We choose the first value due to the 
lower dispersion of results. It is worth noting, that
\Vm = 0.5 km/s seems to be 
reasonable from the point of view that the lower opacity photospheres of metal poor stars allow
spectra to probe  deeper, i.e. into  higher pressure layers where we cannot expect high 
velocity motions/large \Vm. 

\subsubsection{Fit to sodium resonance doublet}

Sodium lines are of special interest as they form 
 strong features in the spectra of both \ga and \gb.
 In the primary spectrum we see the strong lines of the  
resonance doublet of Na I at 5889.95 and 5895.92 \AA. 
Their intensities depend on 
\Tef, \logg, \Vm and the sodium abundance. In the bottom panel of  Fig \ref{_vt} we show the fit
of our synthetic spectra  to the Na I doublet 
 line profiles observed in \ga spectrum. 
Both lines are better fitted  with
the model atmosphere \Tef = 4625, log g =4.5 with N(Na) = -7.34, i.e,
we obtain [Na/Fe] = -0.2 for the primary.

\subsubsection{Fit to Ti II lines in the observed spectrum of \ga \label{_tii}}

The abundance ratio of ionized species of the same 
 element,
usually Fe I/Fe II, is known 
to change with \logg for a fixed \Tef 
 and therefore it is often used to verify the selection of \logg,
 serving to constrain the global solution.
 However, our primary \ga is metal deficient and a relatively cool star.
 In comparison with the Sun, it is difficult to locate 
 appropriate lines of Fe II. We found a few features consisting of weak and blended Fe II lines, 
 but results of their fit was not confident due to the 
 low quality of the observed spectrum.

\begin{figure*}
\centering
\includegraphics[width=\columnwidth, angle=0]{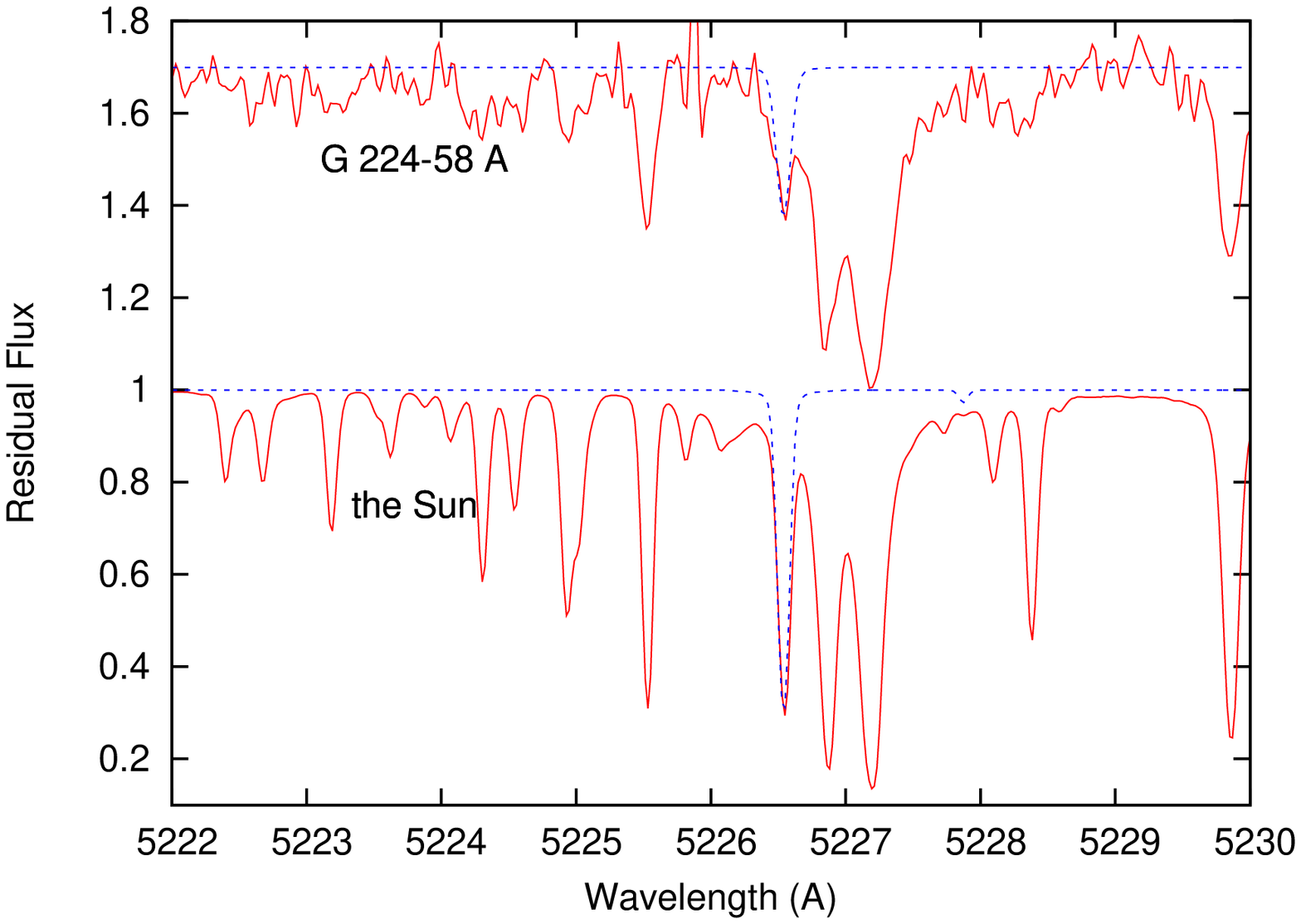}
\includegraphics[width=\columnwidth, angle=0]{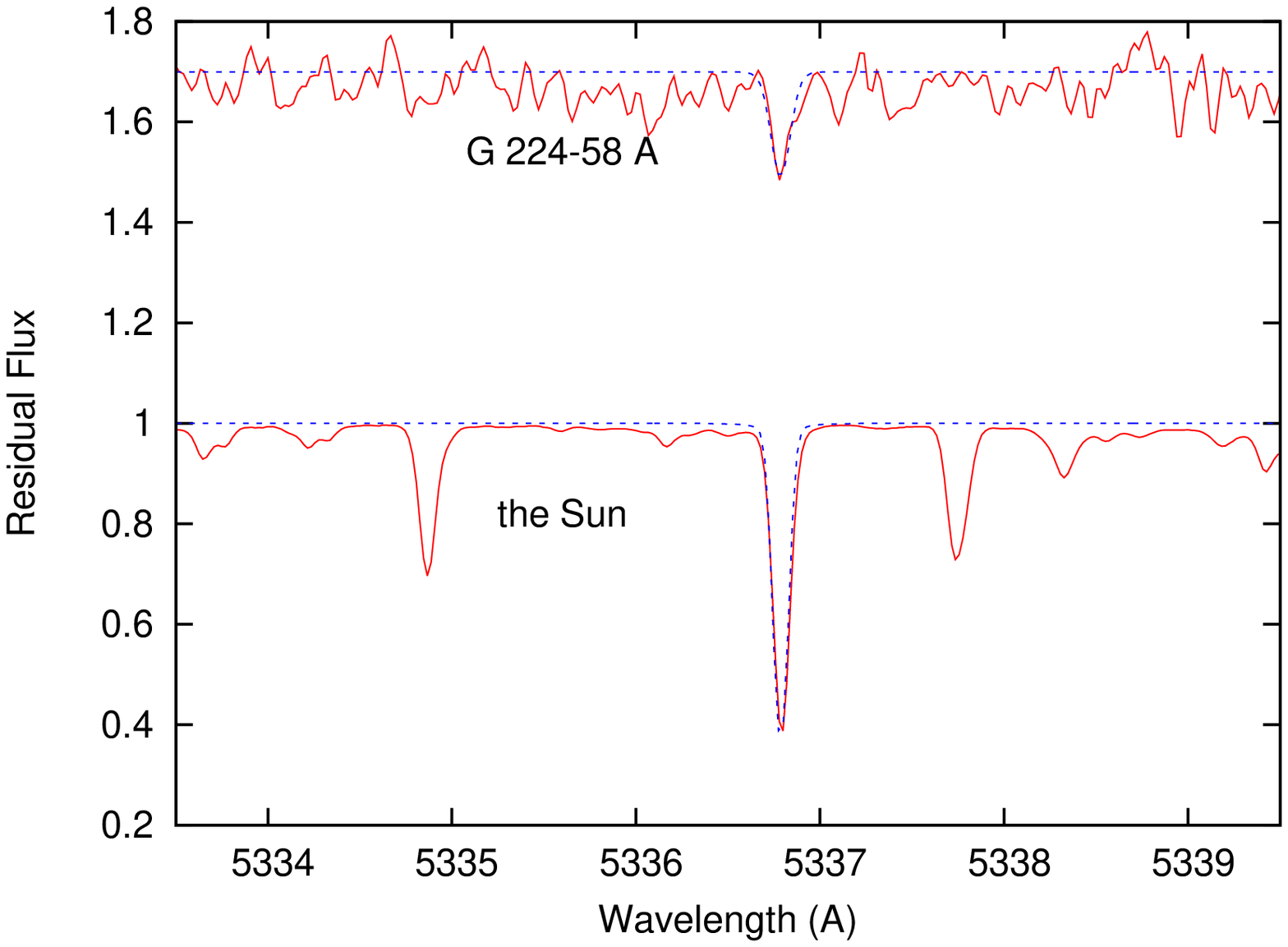}
\caption{Fit of our theoretical Ti II  lines to the 
observed spectra of the \ga and the Sun. In the theoretical 
spectra only absorption by Ti II is included. Spectra of \ga are shifted 
vertically to simplify the plots.}
\label{_ti}
\end{figure*}

On the other hand, due to the lower ionization potential of the 
neutral titanium $(E = 6.83 eV)$
 in comparison to Fe I $(E = 7.90 eV)$, the lines of Ti II should be stronger
 in spectra of comparatively cooler stars, i.e., 
more useful for abundance analysis. 
  Therefore we used lines of Ti I and II to make the analysis instead.
 Indeed, we found lines of Ti II which are fit
 well in our synthetic spectra and list in Table \ref{_TTiII}.
 To be more confident, we also looked at these lines in spectrum of the 
Sun finding
 a good agreement between the lines in model atmosphere 5777/4.44/0 
from \cite{pavl03}
 and observations from the \cite{kuru84} atlas, see Fig. \ref{_ti}.
 The good agreement 
between abundances found using Ti I vs Ti II 
 gives additional independent
 support in our choice for the \ga  model atmosphere parameters.

\subsubsection{Other elements}

\begin{table}
\caption{Fitted Ti II  lines in spectra of \ga and the Sun, 
spectroscopic data were taken from 
VALD3.}
\label{_TTiII}
\begin{center}
\begin{tabular}{ccc}
\hline
\hline
Wavelength, \AA &    gf & E$''(eV)$          \\       
\hline
          5129.16 &    4.571E-02  &  1.892         \\
          5226.54 &    5.495E-02  &  1.566         \\
          5336.79 &    2.512E-02  &  1.582         \\
          5381.02 &    1.072E-02  &  1.566         \\
          5418.77 &    7.413E-03  &  1.582         \\
\hline
\end{tabular}
\end{center}
\end{table}

\begin{table*}
\caption{Abundances of elements determined from fits to \ga hires spectrum. $l$ and $L_{total}$ are a number of 
accounted lines and the total number of lines in pre-selected line list $L$, respectively. Sensitivity of abundances on the changes of 
parameters of model atmosphere with respect to the model 4625/4.5/-1.5$^*$, \Vm= 0.5 km/s model is shown in the latter columns.}
\label{_tS}
\begin{center}
\begin{tabular}{ccccccccc}
\hline
\hline
Elem & $z$&  log N(x)& [X/H]&\Vsini& $l/L_{total}$ &$\Delta$ \Vm = +0.5 km/s & $\Delta$ \logg = 0.5 & $\Delta$ \Tef = -100 K \\
 \hline
Mg & 12 & -5.972 & -1.512   &   2.33 $\pm$   0.42& 6/7  & 0.003                &       -0.007          &          0.087 \\
Ca & 20 & -7.067 & -1.387   &   2.66 $\pm$   0.14& 16/16&  0.038                &        0.012          &          0.107 \\
Ti & 22 & -8.422 & -1.372   &   2.25 $\pm$   0.21& 14/16&  0.029                &       -0.068          &         0.114 \\
Cr & 24 & -8.254 & -1.884   &   2.83 $\pm$   0.12& 9/9  &  0.089                &        0.033          &          0.167 \\
Mn & 25 & -8.609 & -1.959   &   2.50 $\pm$   0.32& 5/5  &  0.040                &       -0.100          &          0.080 \\
Fe & 26 & -6.290 & -1.920   &   2.50 $\pm$   0.10& 44/44&  0.051                &       -0.094          &          0.090 \\
Ni & 28 & -7.598 & -1.808   &   1.89 $\pm$   0.32& 9/9  &  0.089                &       -0.134          &          0.092 \\
Ba & 56 &-11.780 & -1.870   &   1.88 $\pm$   0.38& 4/4  &  0.125                &       -0.125          &          0.166 \\
\hline
\end{tabular}
\end{center}
\end{table*}

Abundances determined from selected line lists are shown in
the Table \ref{_tS}. In the 6-th column of the table we show the number of 
preselected line lists and number of accounted lines.
In particular we note that that light $\alpha-$elements Mg, Ca, Ti are overabundant in comparison with elements of 
iron peak Cr, Mn, Fe, Ni, Ba. For this analysis we use the weaker  lines,
i.e not ones that we find to be saturated in the observed spectrum.  
Table \ref{_tS} indicates that 
the uncertainties of the abundances do not exceed 0.17 dex for the adopted 
range of parameter variation. 

The found overabundance of Mg, Ca, Ti with
  respect to the Fe shows rather weak
dependence on the choice of \Tef, \logg or \Vm, see Table \ref{_tS}. 
In  general, the overabundance of 
alpha-elements is a well known phenomenon for metal-deficient stars, e.g.,
\cite{maga87, mcwi97, sned04}, or more recently \cite{hans15}. 

 To carry out the process of abundance analysis we fitted the observed profiles of the spectral lines.
This procedure allows us to determine the rotational velocity \Vsini. It is worth
noting that the values of \Vsini in Table 2 should be considered rather as upper limits due to the 
natural 
restrictions provided by the limited quality of our observed spectrum. In some sense \Vsini is used here as 
the adjusting parameter of our fitting procedure. For more a accurate determination of \Vsini we should 
adopt/develop 
more sophisticated models of instrumental broadening and macroturbulence and use them
for fits to the observed spectra of better quality. For now we can claim that \ga is a slowly 
rotating (\Vsini $<$ 3 km/s) star, this agrees well with its status of old halo star.

\subsection{LIRIS infrared spectrum of \ga}

\begin{figure*}
\centering
\includegraphics[width=75mm, angle=0]{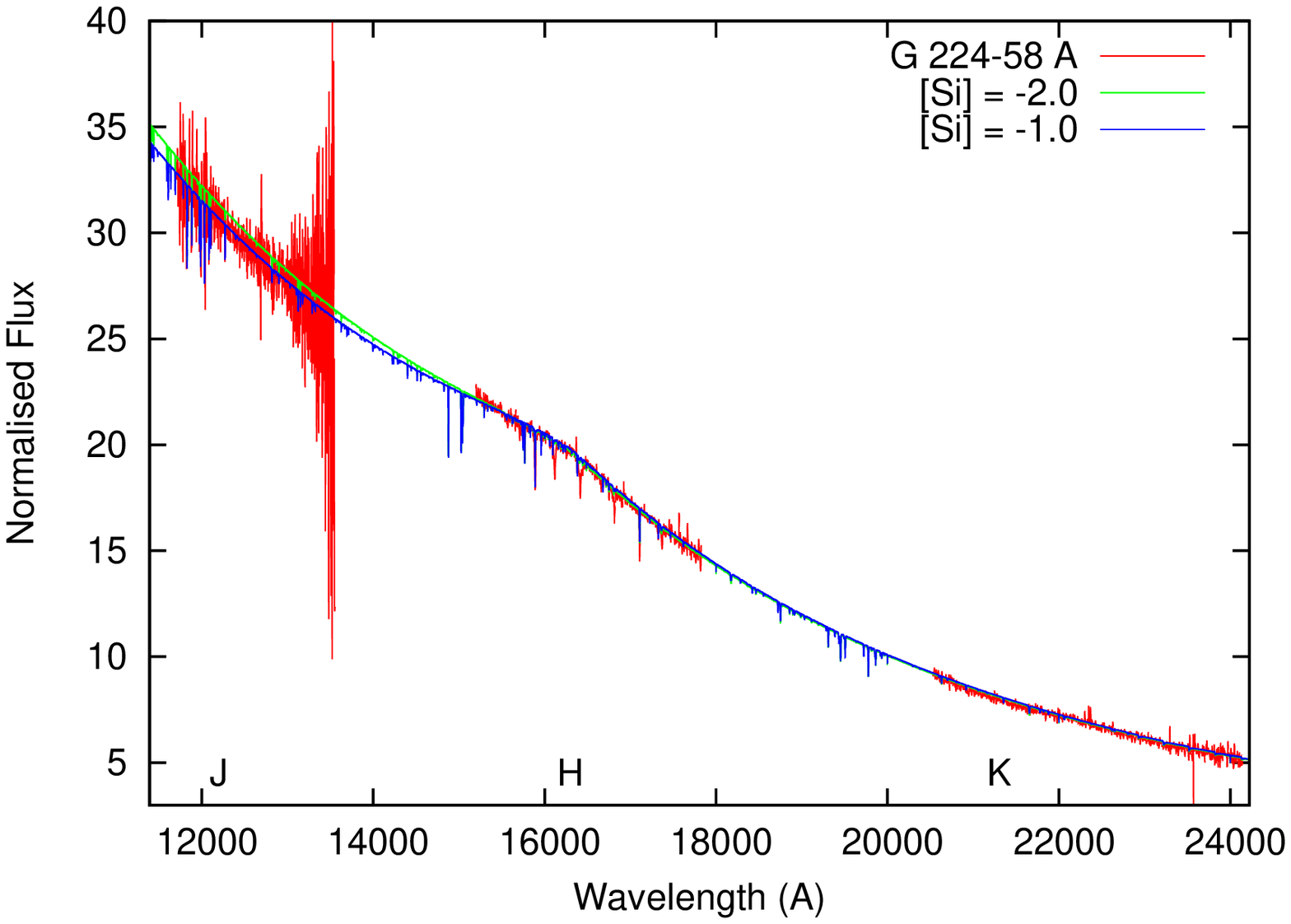}
\includegraphics[width=75mm, angle=0]{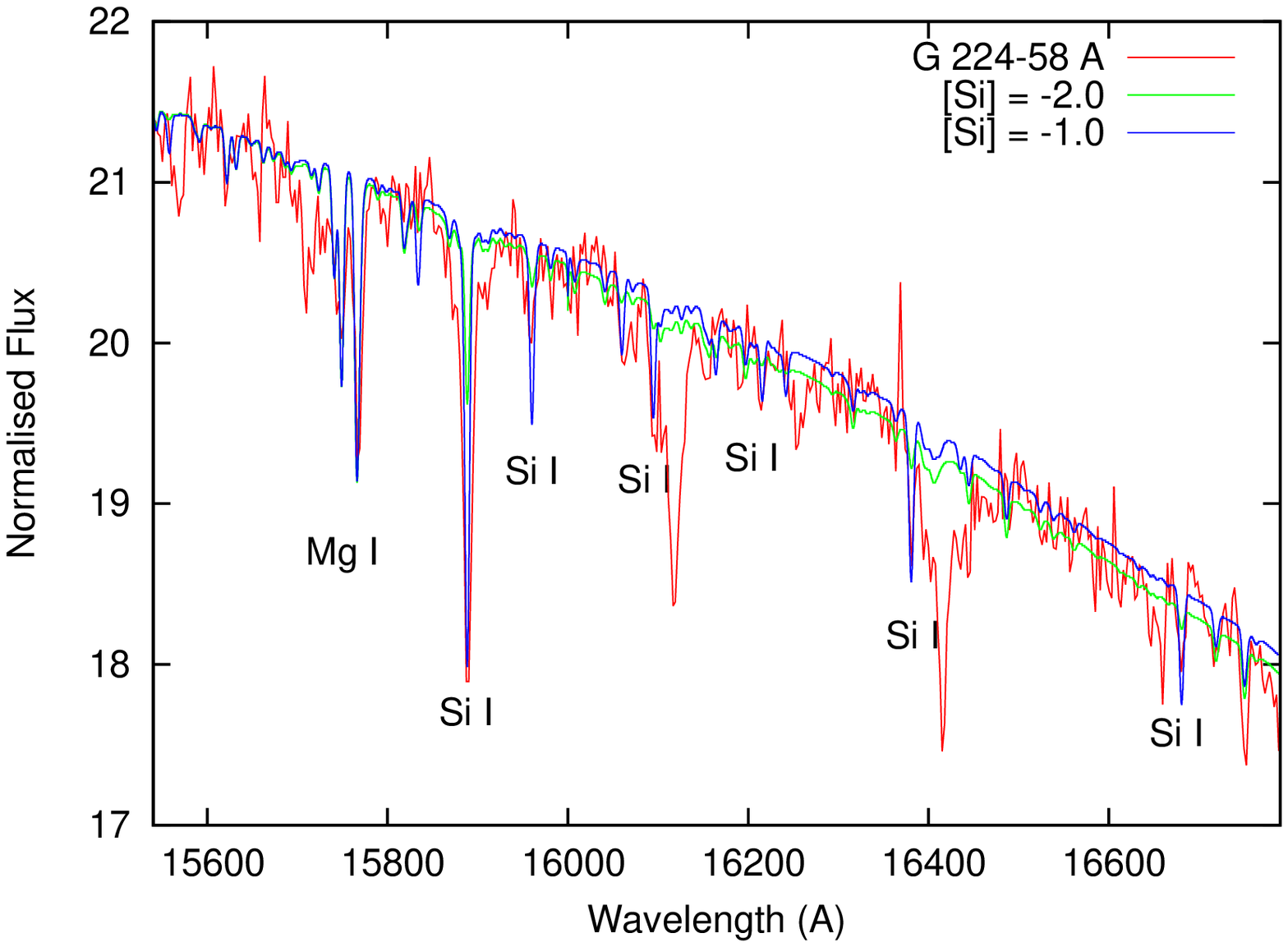}
\includegraphics[width=75mm, angle=0]{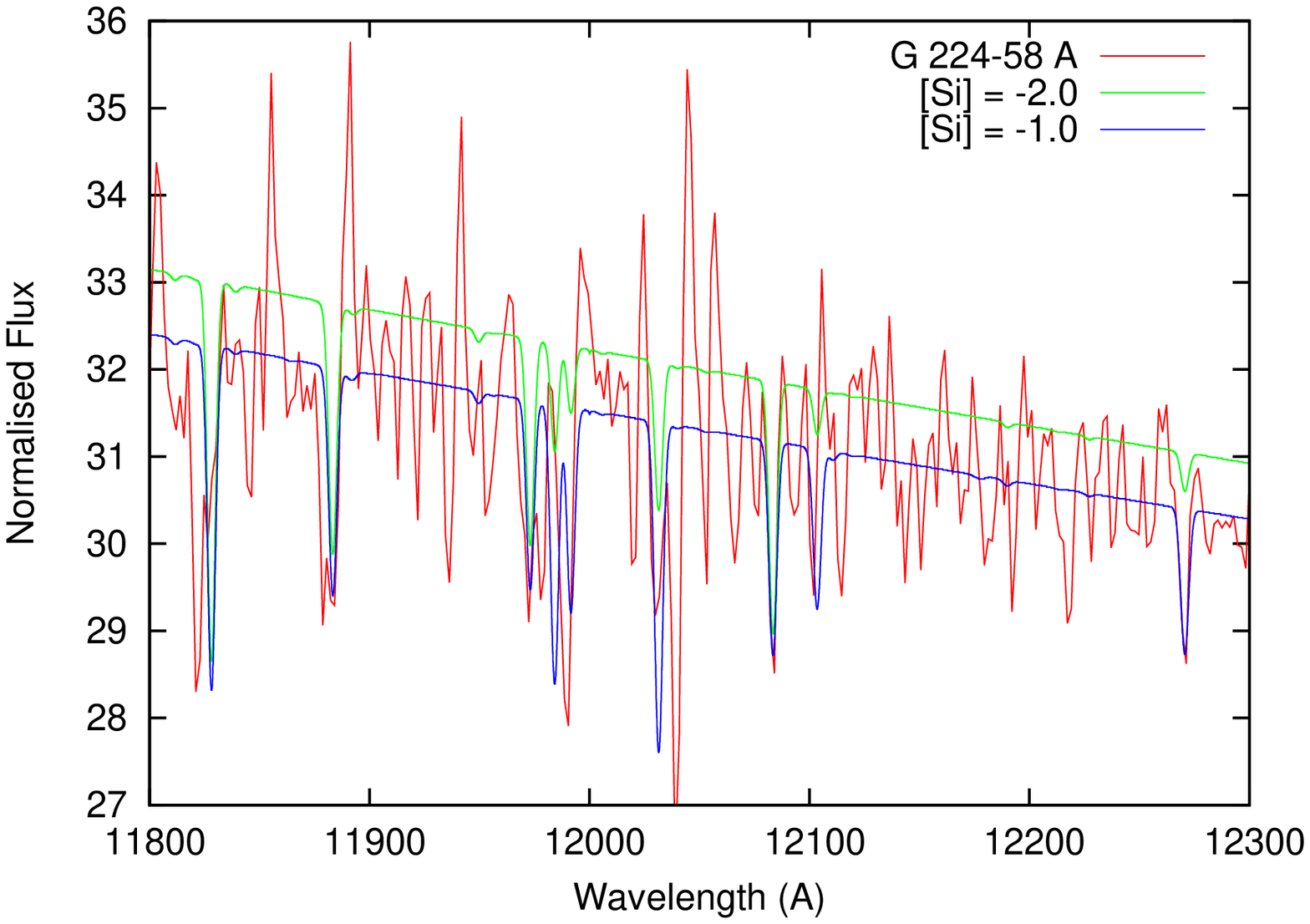}
\includegraphics[width=75mm, angle=0]{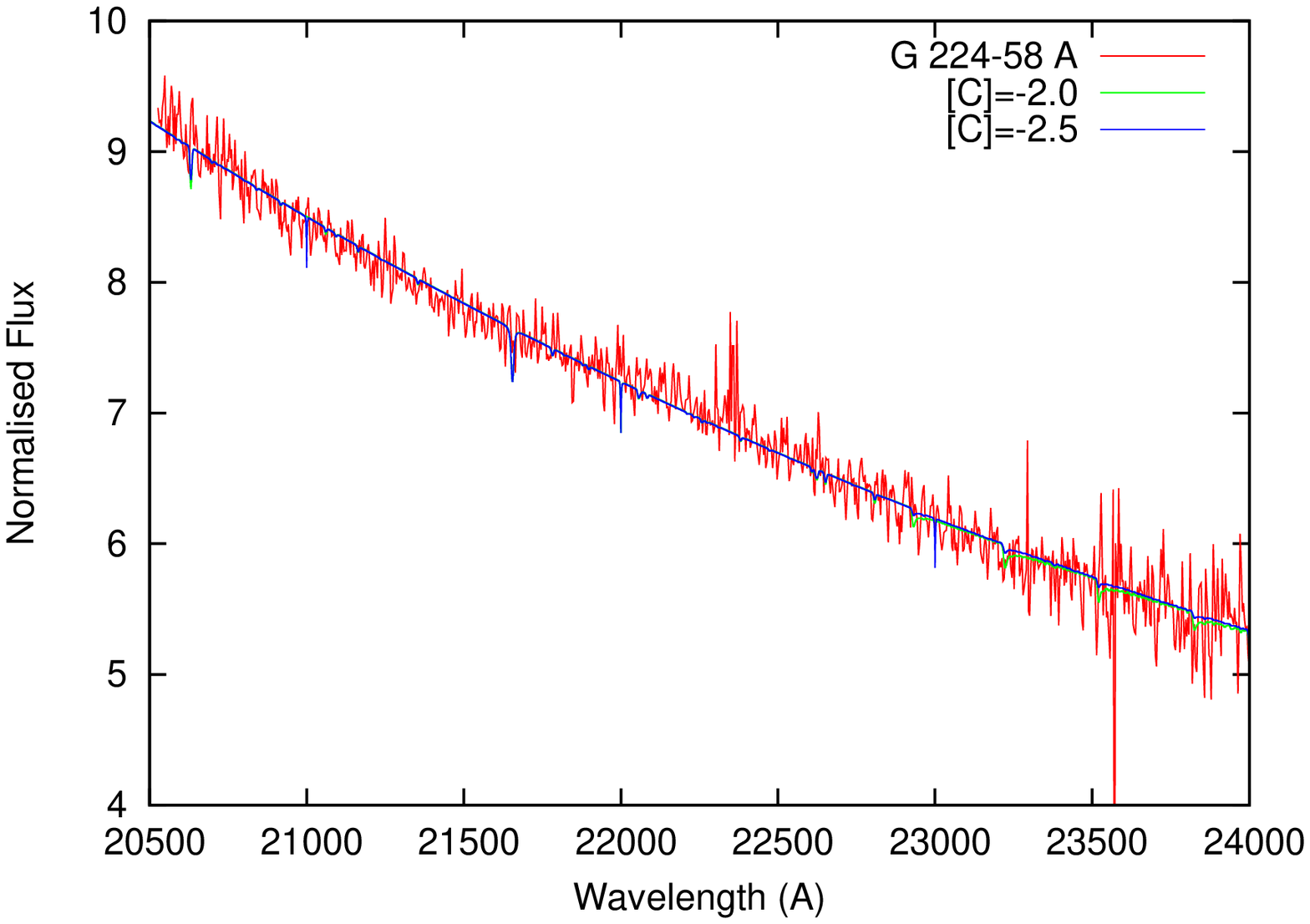}
\caption{Fit to \ga LIRIS  infrared spectrum (top left). 
Three others plots show fits
to the observed  $H$  (top right), $J$ (bottom left), $K$ (bottom right) 
spectral regions in larger scale}
\label{_Aliris}
\end{figure*}

Fits to LIRIS low resolution (R=2500) spectra observed in $H, J, K$ spectral
ranges are shown in Fig. \ref{_Aliris}. The Figure
 shows the observed spectra, where we overplotted the theoretical spectra
 (blue and green lines), for two different Silicon abundances. The
 theoretical spectra were computed by using the optical \ga 
 spectrum (see section \ref{_optA}). 

 Despite the comparatively low resolution, we are able to identify
  a few Si I lines in the observed spectra.
   In the Fig. \ref{_Aliris} we show two cases, with [Si] =--1.0 and --2.0.
It is apparent that Si=--1.0 is a better fit and thus Si I is overabundant.
 Si (Z=14) belongs to the alpha-element group and as discussed earlier
 its overabundance is expected.
Here we can 
claim only an estimation of Si overabundance, i.e. [Si/H] $\sim$ --1.1 
$\pm$ 0.3 dex based on an average of the line fits.
Here we note that changes of Si abundance by factor 10 affects the general 
shape of the observed SEDs (Fig \ref{_Aliris}). Si I is 
 an important donor of free electrons in cool atmospheres, therefore both
 model atmospheres and computed spectra show dependence on its abundance.
 Unfortunately, the absorption lines of Si I are too weak to discern in our optical
 spectrum of \ga lower limit.

In the $H$ spectral region we see a strong Mg I line at 15754 \AA. We obtained
a fit to this line with the Mg abundance obtained from the analysis of optical 
spectrum. In fact, this provides independent confirmation of our results in Table 2.

In the $K$ spectral region we can discern the presence of the CO overtone 
bands at the expected level but not sufficiently well to determine the carbon abundance. 


\section{\gb spectrum analysis\label{_gb}}
\subsection{Procedure}

\gb is a  late spectral type dwarf.
In general we will assume that, as 
a binary the B component has the same properties of 
 age and composition that we have found for A component.
 However, we perform a different approach to metallicity determination.

First, we generated a grid of the LTE synthetic spectra 
following the procedure outlined in \cite{pavl97}. We started with the
NextGen model atmospheres of effective temperatures 2800, 2900, 3000, 3100, 3200,
3300, 3400 K, gravities  of log g = 4.0, 4.5, 5.0, 5.5 and 
metallicities [Fe/H] = --1.5 and 2.0. We replaced some abundances
by the values obtained for the \ga  with parameters \Tef=4625 K, \logg=4.5. 
In our computations 
the line lists of VO and CaH were taken from Kurucz's
website\footnote{kurucz.harvard.edu}, for more details see
\cite{pavl14}. The CrH and FeH line lists were computed by
\cite{burr02} and \cite{duli03}, respectively. We upgraded the TiO
line lists of \cite{plez98} with the new version available on his
website\footnote{http://www.pages-perso-bertrand-plez.univ-montp2.fr/}.
Infrared spectra were computed with account of water vapour absorption
lines provided by the EXOMOL group \cite{barb06}.
The spectroscopic data for atomic absorption VALD3 come from the
Vienna Atomic Line Database
\cite{kupk99}\footnote{http://vald.astro.univie.ac.at/~vald/php/vald.php}.
The profiles of the Na{\small{I}} and K{\small{I}} resonance doublets were
computed here in the framework of a quasi-static approach described in
\cite{pavl07} with an upgraded approach from \cite{burr03}.
More details on the technique and procedure are presented in
\cite{pavl06}.

\subsection{Results of the \gb optical spectrum analysis}

\begin{figure*}
\centering
\includegraphics[width=\columnwidth, angle=0]{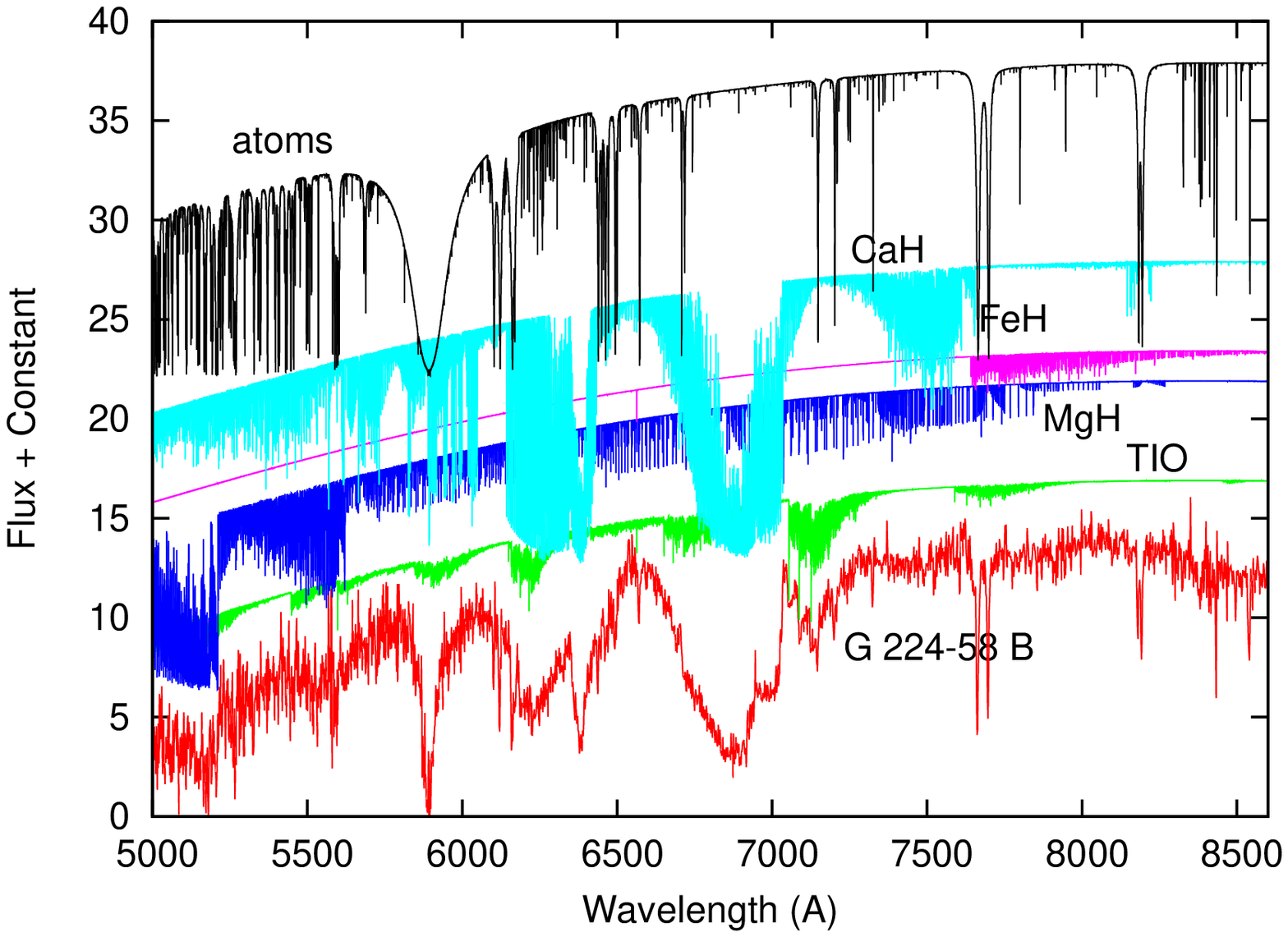}
\includegraphics[width=\columnwidth, angle=0]{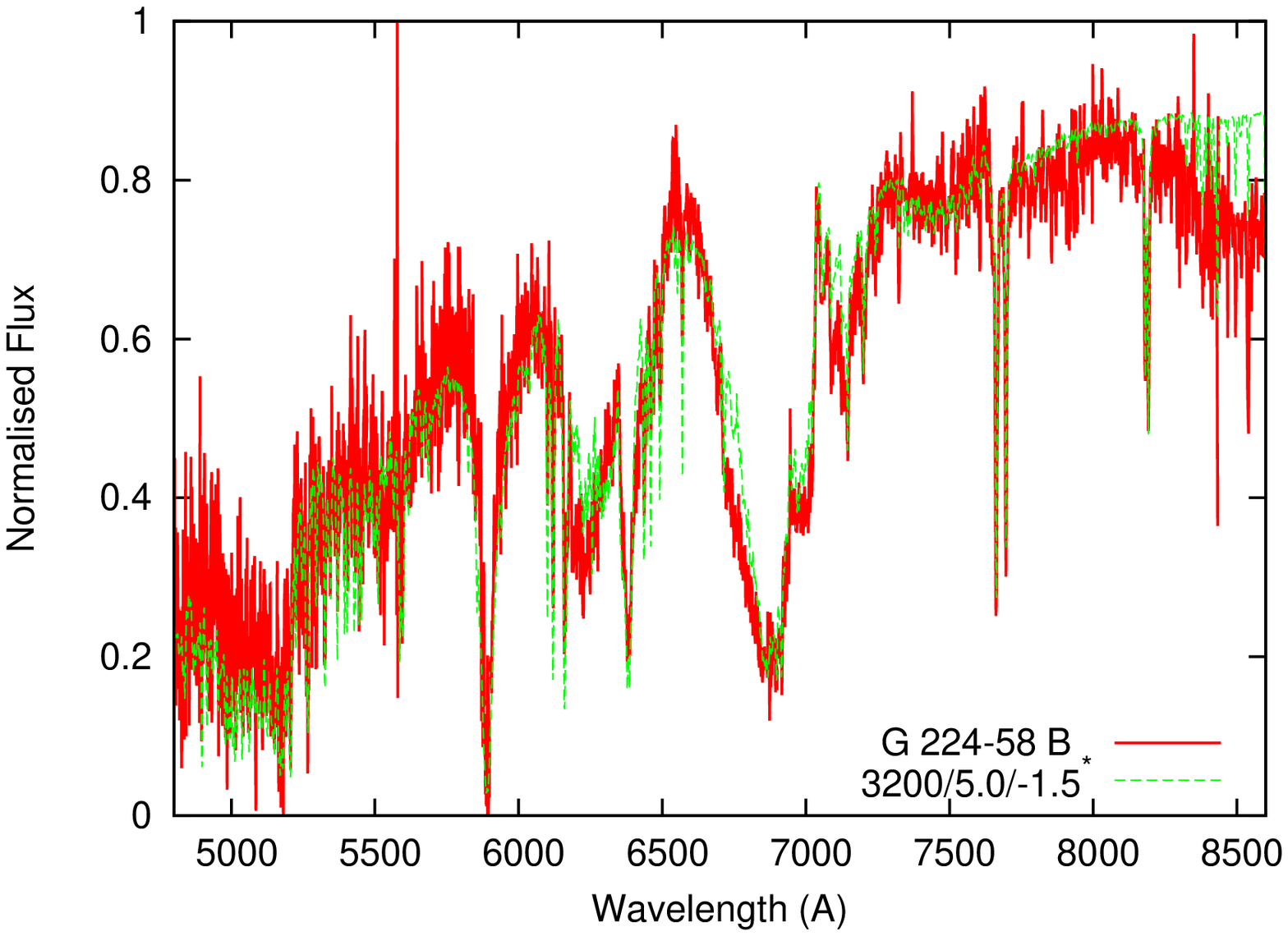}
\caption{Left: identification of the main molecular features formed the
observed optical spectrum of the B component, model atmosphere 3200/5.0/-1.5$^*$
from NextGen grid (\cite{haus99}).
Right: fit 
of our synthetic spectrum computed
with the same model atmosphere to the observed SED of \gb.}
\label{_ident}
\end{figure*}

In Fig. \ref{_ident} we show  a compilation of the
  main absorption features that can be observed  in the spectrum of \gb.
 The main feature is formed by CaH \CaAX band system. 
Strong MgH, TiO band systems
 and K I and Na I resonance lines also form noticeable features in the 
 secondary spectrum. 

 The choice of best fit between the observed and computed spectra
was achieved by minimizing the parameter
$$S(f_{\rm h}) =
   \sum_\nu \left ( F_{\nu} - F_{\nu}^x \right )^2  , $$
where $F_{\nu}$ and $F_{\nu}^x$ are the fluxes in the observed and computed spectra,
respectively, and $f_{\rm h}$ is the normalization factor. 
A similar procedure
was used by \cite{pavl06}.
Fits were made for all synthetic spectra
in our grid.
We use the instrumental  resolution  $R=1500$ for our calculations. 

\begin{figure*}
\centering

\includegraphics[width=75mm, angle=0]{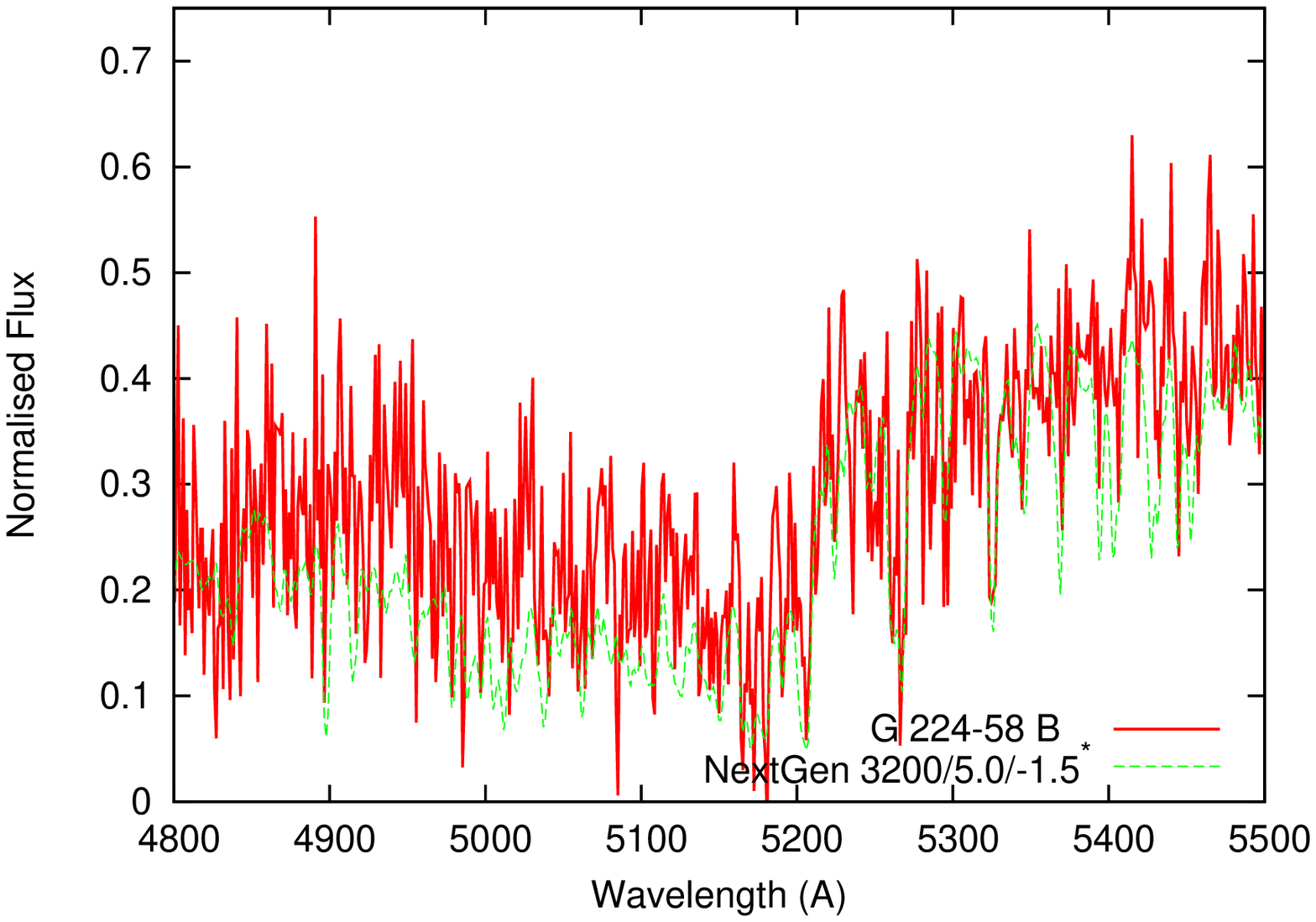}
\includegraphics[width=75mm, angle=0]{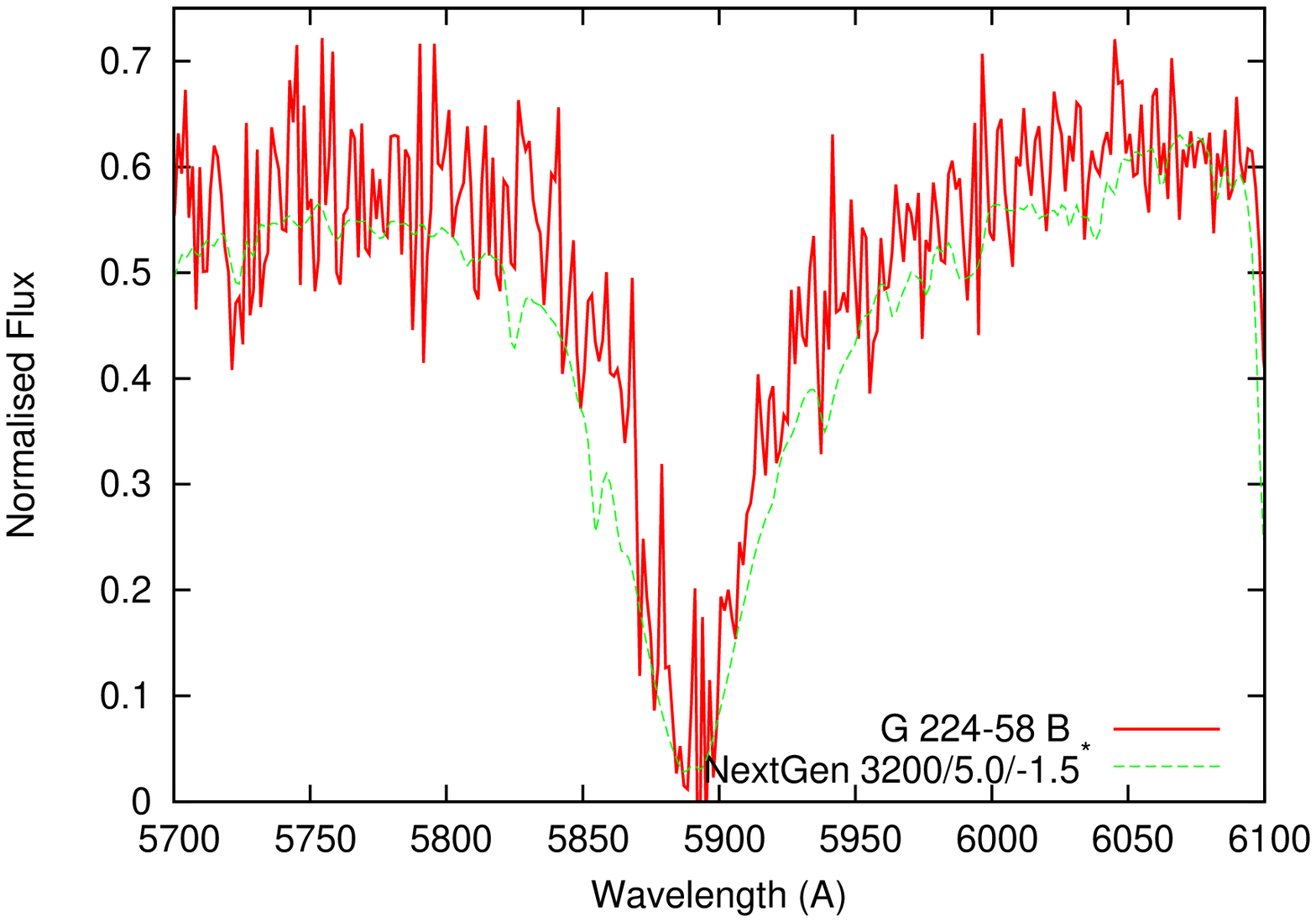}
\includegraphics[width=75mm, angle=0]{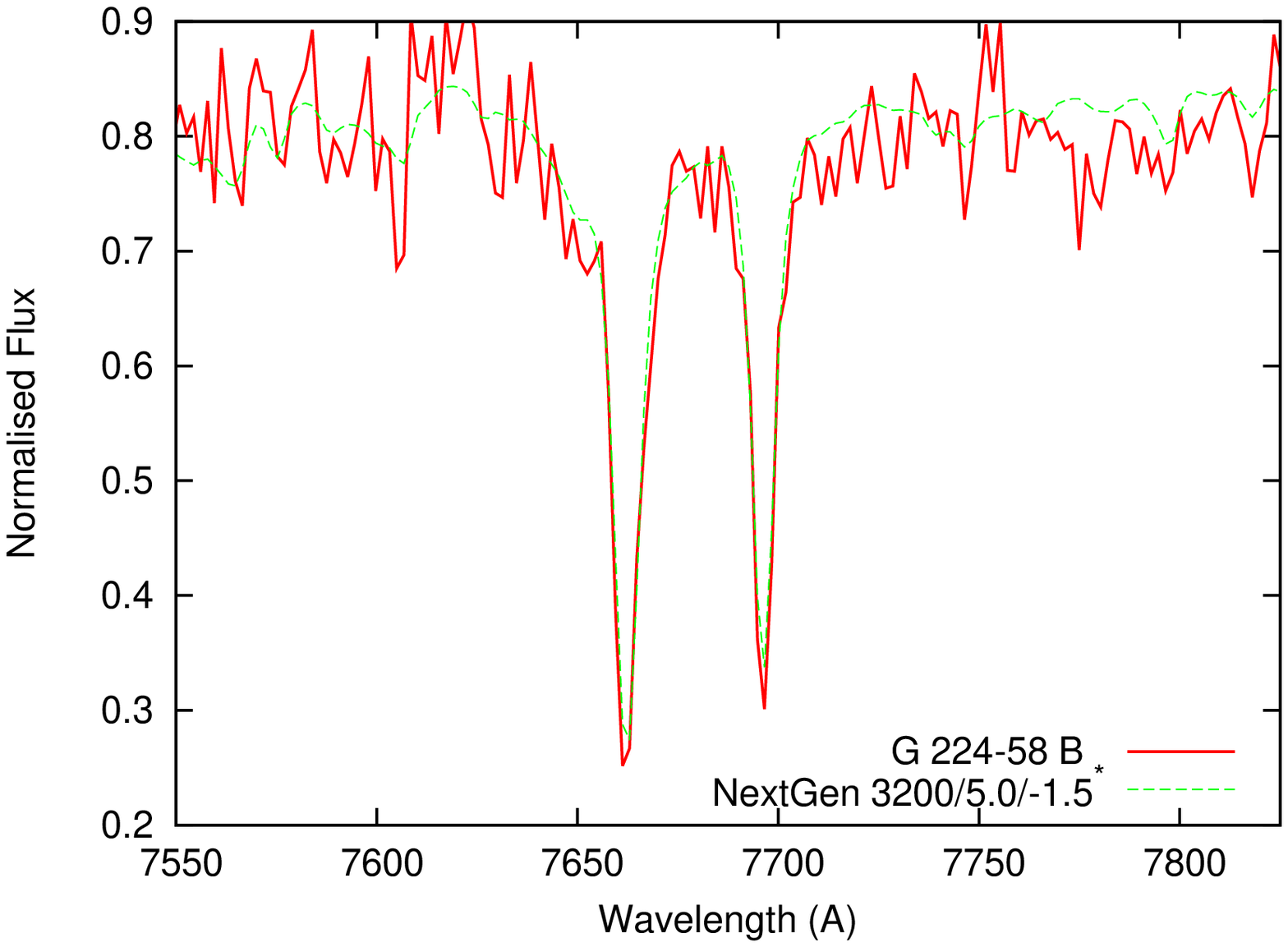}
\includegraphics[width=75mm, angle=0]{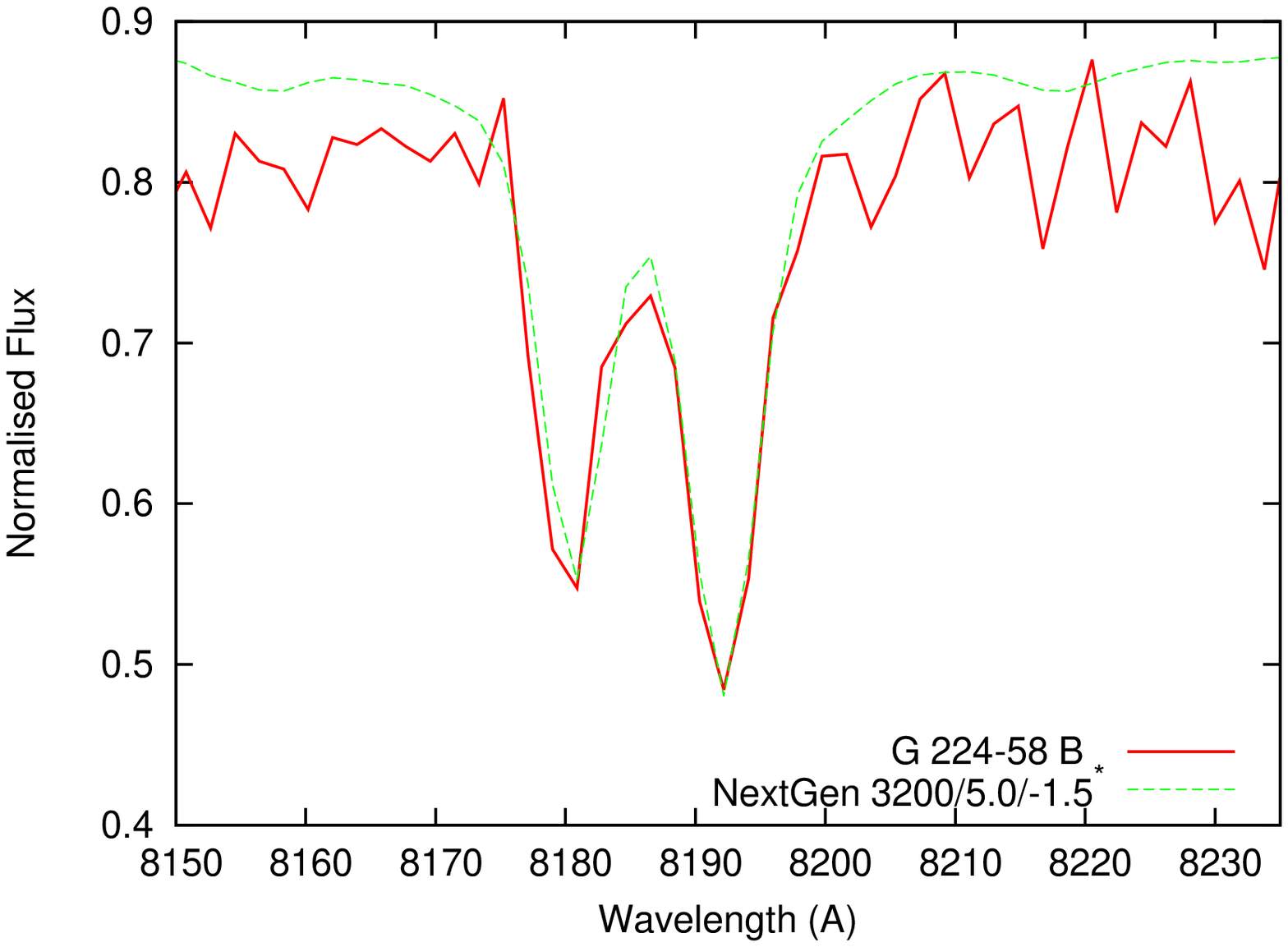}
\caption{Fits of some spectral details in our theoretical spectra 
shown in the right panel of the Fig \ref{_ident} to the observed features in \gb 
spectrum: MgH \AX band system at 5200 \AA (top left), lines of Na I resonance doublet 
(top right), lines of resonance doublet of K I (bottom left), subordinate lines of Na I
(bottom right). Model atmosphere and normalization point are the same as for Fig \ref{_ident}.
}
\label{_FitsB}
\end{figure*}

The best fit of our synthetic spectrum to the \gb observed SED
 is obtained for the 3200/5.0/-1.5$^*$ model atmosphere (see Fig. \ref{_FitsB}). 
Abundance scaling parameter -1.5$^*$ means here that a
 reducing factor -1.5 was used for all abundances in the NextGen
  model atmosphere except for those obtained from the analysis 
 of the primary spectrum. Then, we used ''the best fit'' (as
 defined previously)  to describe the best agreement between
 the computed and observed SED and the profiles of strong lines.
 Fig. \ref{_FitsB} shows the fit to the observed
 SEDs and intensities of the strong sodium and potassium lines, with
 abundances obtained from analysis of \ga spectrum. 

\subsubsection{Strong features in the \gb spectrum}

The notable features in the observed spectrum are formed by MgH \AX band 
system at 5200 \AA, CaH \CaAX and \CaBX
band systems in the wide spectral range, 
as well as TiO $\gamma$ band system at 7200 \AA. We see that these
features in the observed spectrum are well reproduced by our theoretical 
computations with the abundance obtained from \ga analysis. 

Furthermore, in the observed \gb spectrum we see
sodium and potassium resonance doublet lines at 5890 and 7680 \AA,
as well as
sodium triplet lines at 8190 \AA \ known as a good gravity discriminator.
In Fig.\ref{_FitsB} we show our fits to these observed K I and Na I lines. 
To reproduce Na I lines in \gb spectrum we used the sodium abundances determined
from analysis of \ga spectrum. 

The potassium resonance lines are beyond
the spectral range observed for the \ga, and other potassium lines are very weak.
Still we see strong lines of potassium doublet in \gb
observed spectrum. To fit  these lines we used the same abundance
scaling factor, as was found for sodium, i.e. [K/Fe] = --0.2 and obtained 
good agreement between observed and computed profiles. It would be interesting to
verify this result by the comparison of computed and observed profiles of
potassium resonance doublet in the \ga spectrum. 

 \subsubsection{Fe lines}

Since iron lines in \ga spectrum are numerous,
 the  determination of Fe abundance was done with high accuracy.
Ideally a similar analysis would be performed for the \gb, using the 
same FeI line list. However, practically its lower effective temperature, where
  molecular bands dominate the spectrum combined with the lower resolution and 
signal-to-noise make this difficult. The lines are at  the level of the noise. We could use only the Fe I line at 5325.19 \AA.
 We only can claim that weak Fe I lines are present in the observed spectrum and based on Fig \ref{_FeB} find [Fe/H] = --1.7.  

\subsection{Fit to LIRIS spectra of \gb}

The LIRIS infrared spectrum of \gb has  
comparatively low signal-to-noise. Therefore, to reduce the level of noise we
binned the observed spectrum by a factor of 7.
In the following we show the binned spectra to simplify. 

\begin{figure*}
\centering
\includegraphics[width=\columnwidth, angle=0]{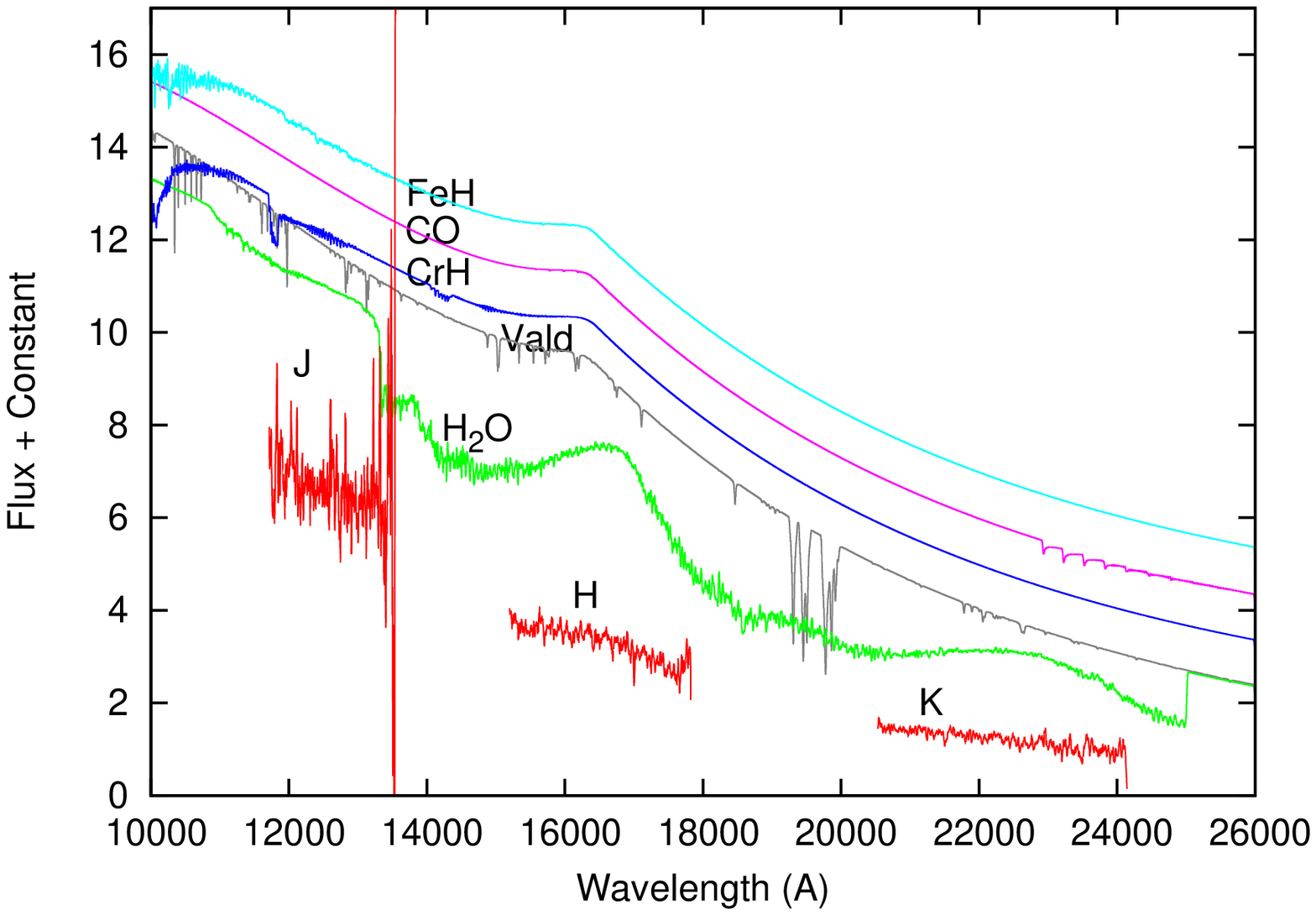}
\includegraphics[width=\columnwidth, angle=0]{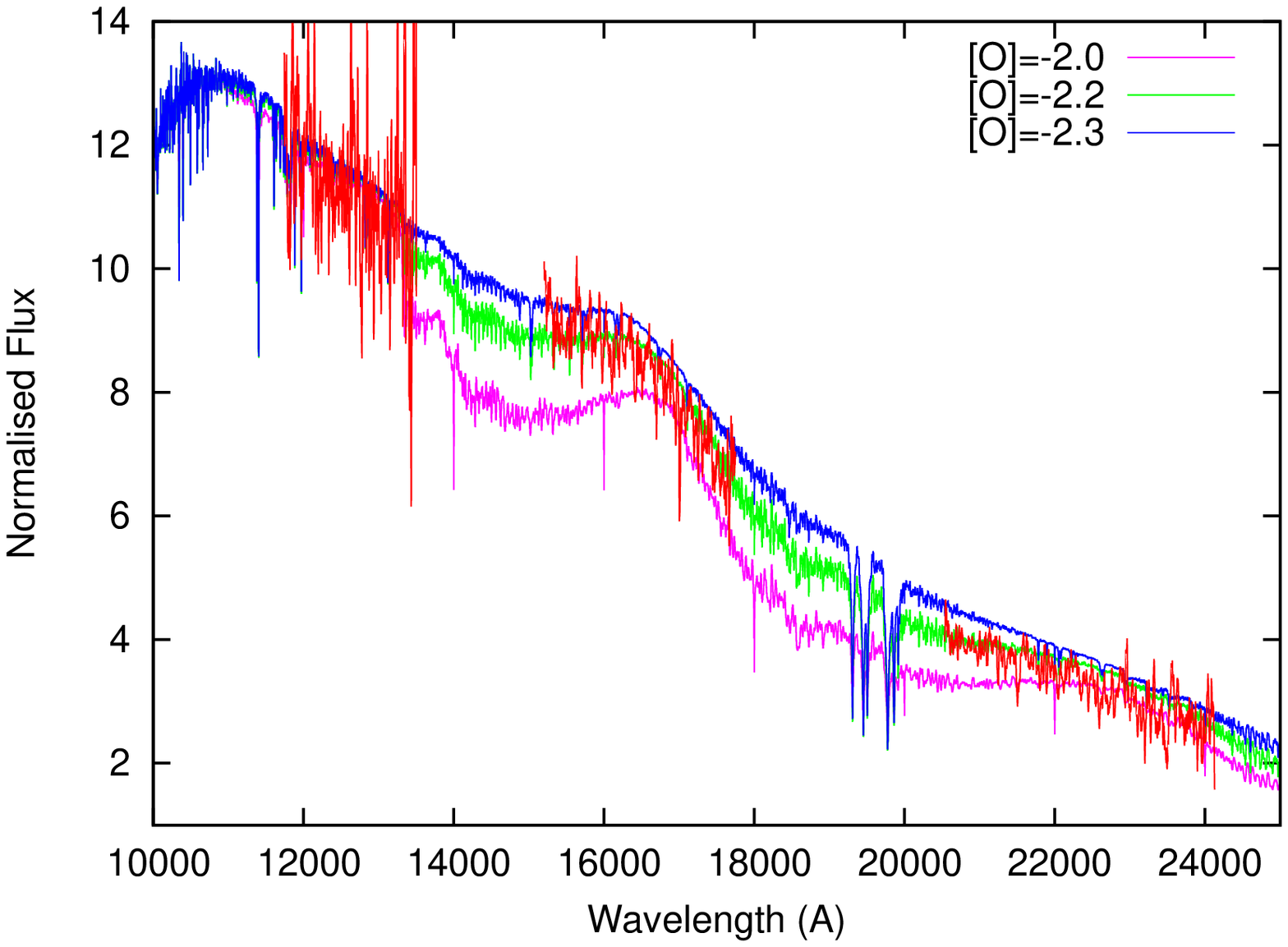}
\caption{Left: identification of the main molecular features formed the
observed by LIRIS infrared spectrum of the B component, 
model atmosphere 3200/5.0/-1.5$^*$.
''Vald'' 
marks the synthetic spectrum computed for atomic line list.
 ''H'', ''J'', ''K'' mark three  spectral ranges
observed by LIRIS.
 Right: fit 
of our synthetic spectra computed
with the same model atmosphere, but different oxygen abundances, to the 
observed SED of \gb.}
\label{_irBid}
\end{figure*}

\begin{figure}
\centering
\includegraphics[width=\columnwidth, angle=0]{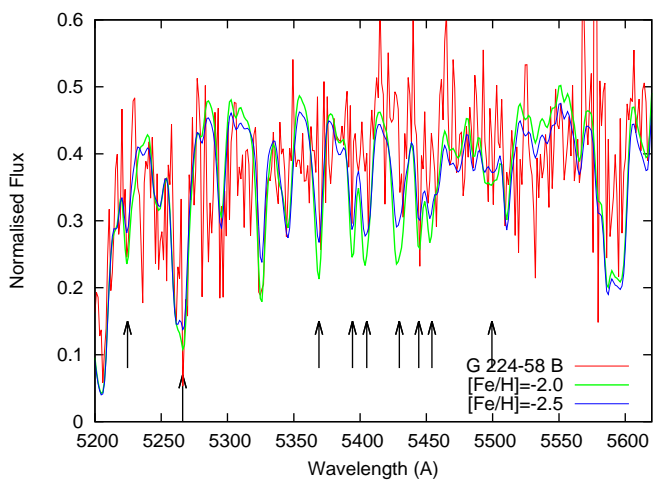}
\caption{Fit of our computed 3200/5.0/-1.5$^*$ model atmosphere 
theoretical spectrum to
the observed \gb spectrum. NIR spectra were smoothed here by 3 pixels for clarify.
Vertical arrows mark positions of Fe I lines in the computed spectrum.}
\label{_FeB}
\end{figure}

Synthetic spectra were computed for the model atmosphere 3200/5.0/-1.5$^*$ with
a 0.05 \AA \ step taking into
 account all known molecular opacity sources.
 First of all we computed the contribution of the molecular bands
 to the opacity in our spectral ranges. 
 Results are shown in the left panel of Fig \ref{_irBid}.
Water vapour absorption dominates across all modelled spectral ranges.
In the most general case absorption by H$_2$O depends on effective temperature,
oxygen abundance and other input parameters. In our work we fixed almost all
of these parameters, so we have a chance to estimate the oxygen abundance.
Indeed, the shape of the computed SEDs across the H band depends on the 
water absorption, with better agreement when we adopt some underabundance
in the atmosphere of the B component, see the right panel of Fig. \ref{_irBid}.

\subsection{Fit of BT-Settl spectra to the optical and infrared SEDs of \gb}

Here we computed synthetic spectra for model atmospheres from the
NextGen grid \citep{haus99} as newer BT-Settl model atmosphere 
are not in the public access.  Nonetheless, fluxes for a fixed grid of
abundances are available and we used them for the comparison
with our observed spectra. 

\begin{figure*}
\centering
\includegraphics[width=\textwidth, angle=0]{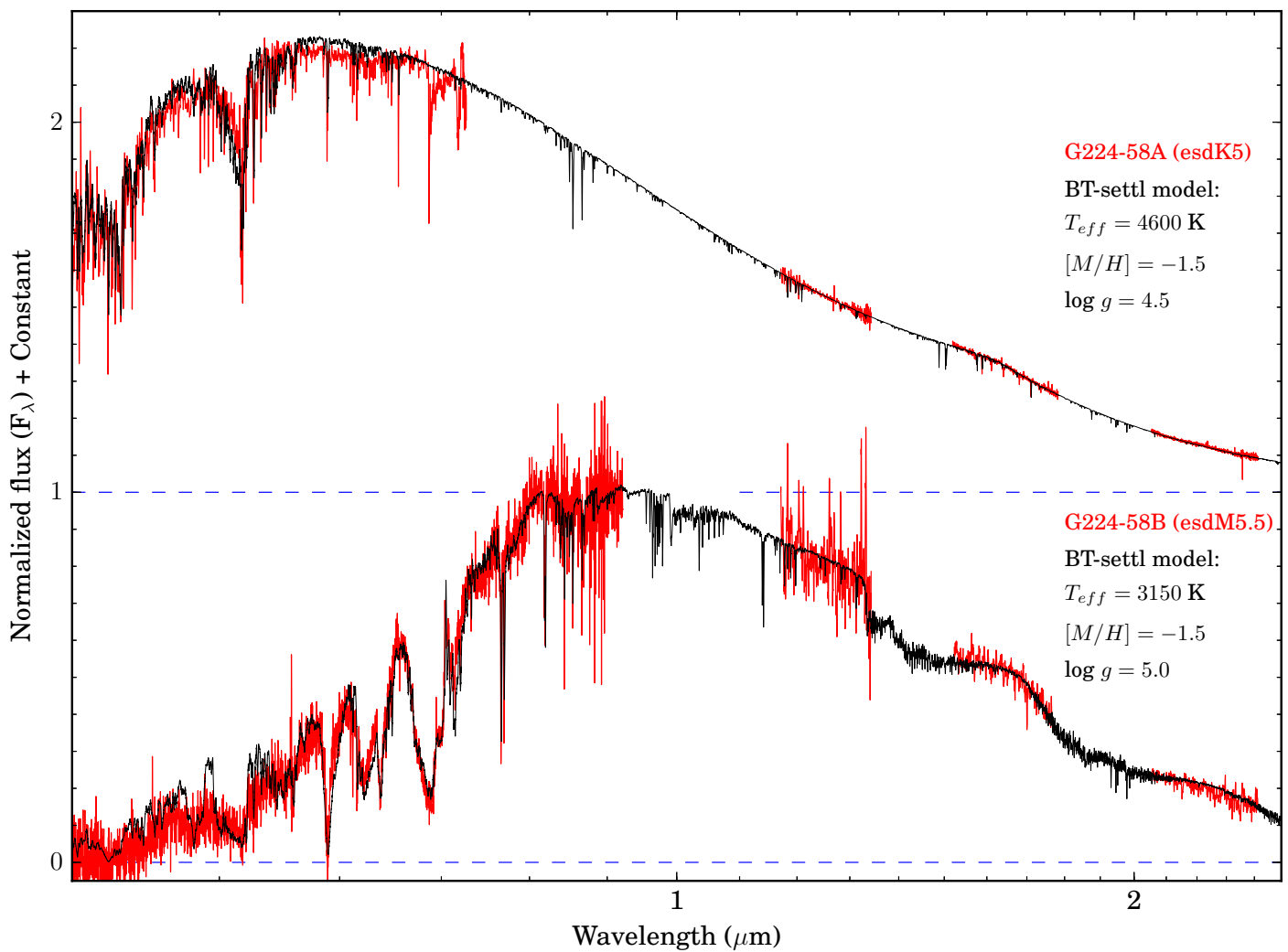}
\caption{Spectra of \gab ~compared to  
BT-Settl  model spectra. Spectra are normalised at 0.82 $\mu$m.}
\label{bt}
\end{figure*}

In Fig. \ref{bt} we provide fits of the BT-Settl spectra 
computed with 3150/5.0/-1.5  model atmosphere to the observed fluxes of \gb. 
Here the metallicity parameter [Fe/H] = -1.5 was used to adjust all abundances.
As can be seen, BT-Settl fluxes  reproduced well the observed SEDs.
Also the Ca, Mg and Ti abundances adopted in BT-Settl model agree well with 
 our values, obtained from the fits to the A component.
 Therefore CaH, MgH and TiO bands are fitted well enough with both our 
 model with modified abundances and an original BT-Settl model. 
 We should remark that
 the NextGen and BT-Settl model atmospheres should be similar. Indeed,
dusty effects at these relatively warm temperatures from the BT-Settl model 
 atmosphere will be very minor, given the case of similarity of the main opacity sources 
 we should get similar results.

\section{Discussions and conclusions\label{_dc}}

 Our final results of abundance determination in atmospheres of \ga and \gb
are shown in Table \ref{_tfin}.
Here we compiled abundances obtained by fits to our spectra across optical and infrared 
spectral ranges, using different approaches to get the best fits to observed
absorption profiles and spectral features. In that way we obtained the fits to 
\ga and \gb spectra using the same abundances.

\begin{table}
\caption{Abundances determined for the atmospheres of \ga and \gb. The errors given for \ga are determined by our minimisation procedure and for \gb are made ``by-eye''.}
\label{_tfin}
\begin{center}
\begin{tabular}{ccccc}
\hline
\hline
Element & $Z$  & log N               & $[X/H]_A$  & $[X/H]_B$       \\
\hline
 O  &  8  & -5.3 & & -2.1 $\pm$ 0.2       \\
 Na  & 11 & -7.87 & -2.1 $\pm$ 0.2 & -2.1 $\pm$ 0.2             \\
 Mg  & 12 & -5.975 & -1.512 $\pm$  0.079  & -1.5 $\pm$ 0.1 \\
 Si  & 14 & -5.59  & & -1.1 $\pm$ 0.3    \\
 K & 19 & -9.0  & & -2.1$\pm$ 0.2      \\ 
Ca & 20 & -7.105 & -1.387 $\pm$0.025 & -1.4 $\pm$  0.2   \\
Ti & 22 & -8.451 & -1.372 $\pm$  0.028   \\
Cr & 24 & -8.343 & -1.884 $\pm$  0.065   \\
Mn & 25 & -8.649 & -1.959 $\pm$  0.055   \\
Fe & 26 & -6.341 & -1.920 $\pm$  0.020 & -1.9 $\pm$ 0.5   \\
Ni & 28 & -7.687 & -1.808  $\pm$  0.050  \\
Ba & 56 & -11.905& -1.870  $\pm$  0.111  \\
\hline
\end{tabular}
\end{center}
\end{table}

\cite{zhan13} provided very strong arguments about the
 binarity of G 224-58 AB, based on the common high proper motion (PM) and radial velocity (RV).
We used their results as a starting point for our investigation. 
 Our study provides confirmation about 
the binary nature of the objects through the consistency of the 
abundance measurements of the A and B components.


 From our analysis of the primary spectrum (see Table \ref{_tS})
we determined [Mg/Fe] = +0.41$\pm$0.05,
[Ca/Fe] = +0.53$\pm$0.05,
[Ti/Fe] = +0.55$\pm$0.05,
i.e. these elements are overabundant relative to iron.
Our finding
 is confirmed by direct modeling of the cool 
secondary dwarf spectra.
 Generally speaking, the overabundance of Ca in atmospheres of metal poor stars
 is well known  \citep[see][and references therein]{maga87, mcwi97, sned04}. 
 In the case of binary systems, or exoplanetary
 systems the situation looks even more intriguing, because here we have the impact
 of a few very different processes: early epoch nucleosynthesis, separation of 
 heavy elements in the binary system process formation, etc.
An interesting aspect of studying halo systems including a cool component
 like G~224-58~AB is that it provides an opportunity
for the abundance determination of some elements, e.g. K and Na.

 Generally speaking, for [Fe/H] $>$ -1.0 there is not
 much variation in abundance ratios for different elements at any given
[Fe/H] for single (non-binary) main sequence stars, so that knowing
[Fe/H] does give us a good idea what its abundances are for other
elements, e.g., figures 9, 10, and 11 in \cite{redd03}.
Besides, for lower metallicity stars there is more scatter in the [X/Fe] ratio
 for some elements for a given [Fe/H], which may affect the stellar
atmosphere, e.g. two stars with [Fe/H] = -1.5 but with very different
alpha element abundances will have different line opacity.

The essential result of our \ga spectrum analysis is the conclusion 
that at least for the case of metal deficient stars with [Fe/H] $<$ -1.0 
we cannot use one parameter, the metallicity, 
to describe the behaviour of individual abundances. This conclusion 
agrees well with \cite{wool09} results and demonstrates
the importance and necessity of abundance analysis 
of binaries like G 224-58 AB. 
 Comparison of our results for Ti with \cite{wool09} abundances  
for esdM  shows at least a good qualitative 
agreement. For the most metal poor esdM LP 251-35 (3580/5.0/-1.96) 
they obtained [Ti/Fe] = 0.45. Interestingly, these authors 
 used K I and Ca II lines to determine the parameters of the atmosphere of the stars. 
Here we have obtained an overabundance in calcium and
 under abundance of potassium (and sodium) in the atmosphere of G~224-58.
If the overall abundances of the extremely metal-poor
M-dwarfs are similar,
their results may be affected by abundance uncertainties.

\cite{lepi07} defined a metallicity index, $\zeta_{\rm TiO/CaH}$, which is formed 
by a combination of the spectral indices of TiO5, CaH2, and CaH3. They classified M subdwarfs into 
three metal classes: subdwarf (sdM; $0.5 < \zeta_{\rm TiO/CaH} < 0.825$), extreme subdwarf 
(esdM; $0.2 < \zeta_{\rm TiO/CaH}  < 0.5$ ) and ultra 
subdwarf (usdM; $\zeta_{\rm TiO/CaH} < 0.2$). Metallicity measurements of M subdwarfs have been conducted 
with high resolution spectra  \citep{wool06,wool09,rajp14}. The relationship between the
 $\zeta_{\rm TiO/CaH}$ and [Fe/H] has been studied  \citep{wool09,mann13,rajp14}.  
  We calculated the metallicity index $\zeta_{\rm TiO/CaH}$ = 0.22 
  for \gb to compare to previous works. 

\begin{figure}
\centering
\includegraphics[width=\columnwidth, angle=0]{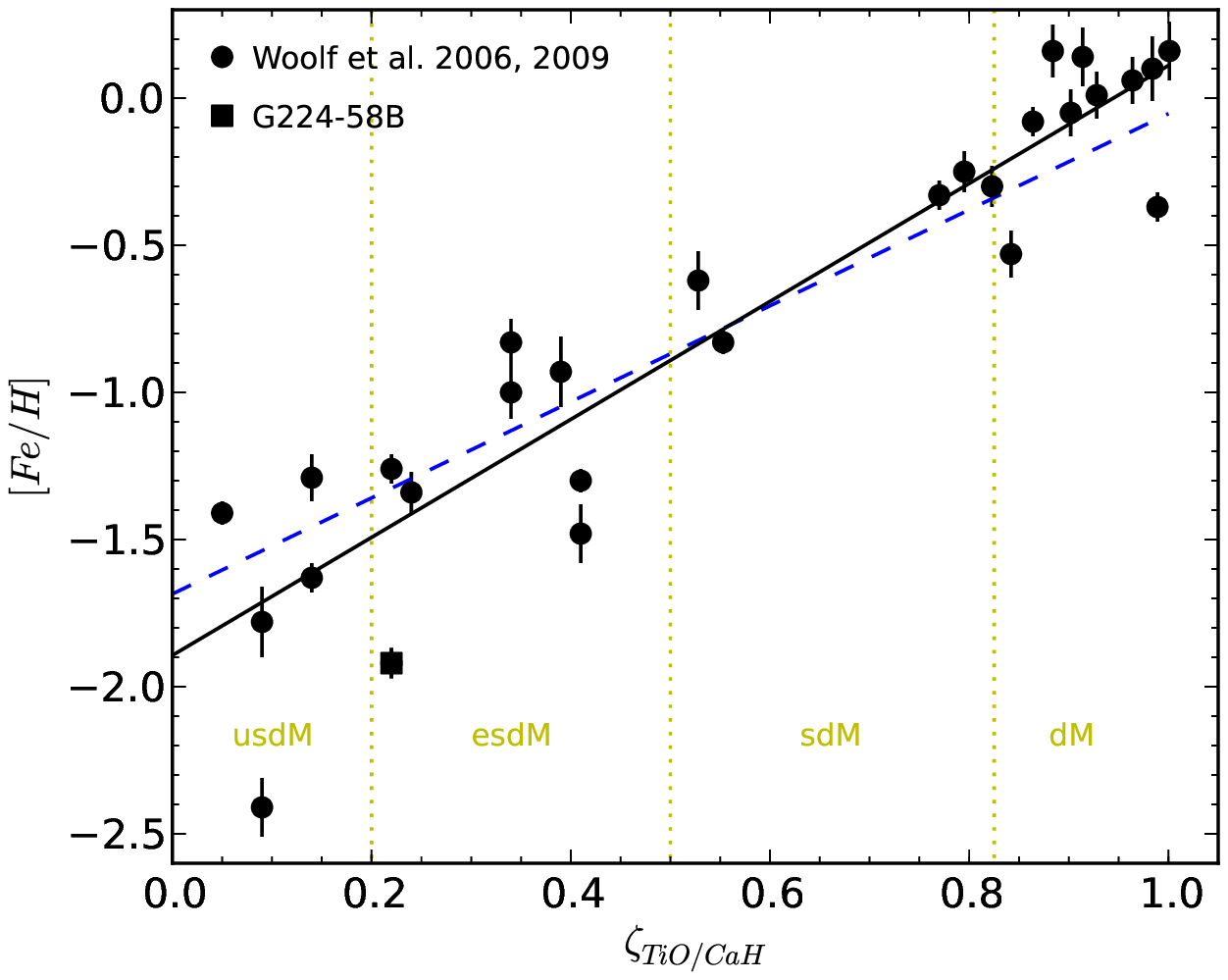}
\caption{Metallicity parameter $\zeta_{\rm TiO/CaH}$ vs. [Fe/H] for \gb and  M subdwarfs from 
\cite{wool06}, \cite{wool09}. 
The blue dashed line represents the fitting from \cite{wool09}. 
The black solid line represents our fitting of \gb and data from \cite{wool06,wool09}. 
The fit can be  described with  $ [Fe/H]  = 2.004 \times  \zeta_{\rm TiO/CaH} - 1.894$. 
Red dotted lines shows the boundaries of metallicity classes of M dwarfs from \cite{lepi07}. }
\label{fe-h}
\end{figure}

Figure \ref{fe-h} shows the $\zeta_{\rm TiO/CaH}$ and [Fe/H] of \gb and 
M subdwarfs from \cite{wool09}.
A blue dashed line in Figure \ref{fe-h} 
represents the correlation of $\zeta_{\rm TiO/CaH}$ and [Fe/H] for late-type K 
and early-type M subdwarfs (3500-4000 K)  \citep{wool09}.   
A black solid line represents our fitting of \gb and M subdwarfs 
from \cite{wool09}, which 
can be described with  [Fe/H]  = 2.004 $\times  \zeta_{\rm TiO/CaH} - 1.894$. 
We excluded objects from \cite{rajp14} in our fitting because to reduce the dispersion
 of results in Fig. \ref{fe-h} would require a shift of [Fe/H] values from \cite{rajp14} by
  a factor of --0.3 to  --0.5 downward. This could arise from the overabundance of Ti and Ca and 
  their inclusion in the [Fe/H] values they compute. 

 It is worth noting that:
\begin{itemize}
\item \cite{rajp14} measure [Fe/H]
with only iron lines,  however,
for three M dwarfs shown in fig. 11 of \cite{rajp14}
 they find $\zeta_{\rm TiO/CaH} >$ 0.825 for
 metallicities from $-$0.5
 to $-$1.0 and so these values are significantly discrepant for the $\zeta_{\rm TiO/CaH}$ parameter.
We note their fit to the sdM9.5 object in a small region of optical
spectrum gave [Fe/H] = -1.1 and \Tef = 3000K. From the fit to the 0.7-2.5
\um NIR spectrum of the same object with same model, we got 
[Fe/H] = -1.6 and \Tef = 2700K. We note that an analysis of a narrow spectral 
region can be significantly impacted by relatively different pseudocontinuums 
formed by molecular bands in the optical and the infrared. The measurement of
 abundance and temperature from atomic line strength can be 
significantly impacted by such differences, e.g., \cite[see][]{pavl95}.
\item Woolf's results of $\zeta_{\rm TiO/CaH}$ determination are 
affected by uncertainties in the adopted 
K, Na, Ca and Mg abundances. 
\item   Our value of iron abundance obtained for the \ga in the analysis explained
 in previous sections is  [Fe/H] = -1.920; 
only one M subdwarf from \cite{wool09} has lower [Fe/H] than \gb but was excluded in their fitting. 

\end{itemize}


For the case of halo dwarfs of later spectral classes, i.e. esdM dwarfs,  
the use of one parameter of metallicity may seriously impact results. 
And, this is why we need binaries to calibrate spectral indices vs. abundance 
dependence for  M subdwarfs of different populations. 

 In general, our fits to the infrared spectra of the \ga and \gb observed 
 with LIRIS provide the independent confirmation of 
 the correctness of our choice of parameters of atmospheres of both components.  
 With the spectrum of \gb, we find evidence of overabundance of Si I and 
 underabundance of oxygen. It is worth nothing these results were obtained
 by fitting to low resolution infrared spectra. In spectra of stars of normal metallicity the results are affected by the
 presence in the spectral analysis of complex blending. However,
 in the case of metal deficient stars these effects are less pronounced.  
 Moreover,  in our case we have not any other chance to estimate 
 oxygen and silicon abundances, except through the modelling of infrared spectra.  
 To analyse weak oxygen and silicon lines in the optical spectrum of A component
 we require the spectral data of  a quality comparable with solar atlas.  
The underabundance found in 
oxygen may be explained as a real small number of atoms in the primary atmosphere due to its
 association in CO when the carbon abundance/presence is comparatively high in the atmosphere.
Our picture could be better tested if were able to do more precise analysis  of the CO
bands at 2.3 micron. 

The abundance inhomogeneities found can be
most easily explained in terms of the nucleosynthesis
  of halo stars at their formation epoch. 
 The low iron abundance, i.e [Fe/H] = -1.92,
 together with large PM determined by \cite{zhan13} allows assignment of G 224-58 AB 
to the halo population.
We detect  in the atmospheres of \gab, overabundances with respect to 
 the iron, not only of Ti, but also of other alpha-element elements.
 The study of similar systems, their physical characteristics as well
 as their multiplicity, is of crucial importance for the investigation of 
 binary and multiple system formation processes and their evolution
  through composition and time, including the formation of 
   still hypothetical exoplanetary systems in the early epochs of Galaxy evolution.

\section{Acknowledgments}
Based on observations made with the Nordic Optical Telescope, operated by the Nordic
 Optical Telescope Scientific Association at the Observatorio del Roque de los Muchachos, 
 La Palma, Spain, of the Instituto de Astrof\' isica de Canarias.
The WHT and its service programme are operated on the island of La Palma by the Isaac Newton Group in the Spanish Observatorio del Roque de los Muchachos of the Instituto de Astrof\' isica de Canarias.
 M.C. G\'alvez-Ortiz acknowledges the financial support of a JAE-Doc CSIC fellowship
 co-funded with the European Social Fund under the
 programme {\em ”Junta para la Ampliaci\'on de Estudios”} and
 the support of the Spanish Ministry of Economy and Competitiveness through the project
 AYA2011-30147-C03-03.
 The authors thank the
compilers of the international databases used
in our study:  SIMBAD (France, Strasbourg),
and VALD (Austria, Vienna) and R. Kurucz and Phoenix group for model atmospheres,
synthetic spectra.
We thank Jose Antonio Acosta Pulido for his helps with the LIRIS data reduction. 
 We thank the anonymous Referee for his/her thorough review and highly appreciate the comments and 
suggestions, which significantly contributed to improving the quality of the publication.

\end{document}